\documentclass[12pt]{article}
\usepackage{amssymb,amsthm,amsfonts}
\usepackage{epsfig}
\usepackage{colordvi}
\def\bea{\begin{eqnarray}}
\def\eea{\end{eqnarray}}
\def\bq{\begin{quote}}
\def\eq{\end{quote}}
\def\nn{\nonumber}

\def\gappeq{\mathrel{\rlap
{\raise.5ex\hbox{$>$}}
{\lower.5ex\hbox{$\sim$}}}}
\def\lappeq{\mathrel{\rlap{\raise.5ex\hbox{$<$}}
{\lower.5ex\hbox{$\sim$}}}}

\newcommand{\beq}{\begin{equation}}
\newcommand{\eeq}{\end{equation}}

\def\varkappa{\omega}

\newcounter{mnotecount}[section]

\begin{document}
\pagestyle{empty}
\begin{flushright}
SACLAY-T06/048\\
\end{flushright}
\vspace*{5mm}
\begin{center}

{\Large \bf Quark-Lepton Unification and Eight-Fold Ambiguity}\\
\vspace*{.3cm}
{\Large \bf in the Left-Right Symmetric Seesaw Mechanism}\\
\vspace*{2cm}

{\bf Pierre~Hosteins, St\'ephane~Lavignac and Carlos A. Savoy}\\
\vspace*{.5cm}

{\it Service de Physique Th\'eorique, CEA-Saclay, F-91191 Gif-sur-Yvette
Cedex, France} \footnote{Laboratoire de la Direction des Sciences de la Mati\`ere
du Commissariat \`a l'Energie Atomique et Unit\'e de Recherche associ\'ee
au CNRS (URA 2306).}\\
\end{center}

\vspace*{1.5cm} 
\centerline{\bf Abstract} 
\vspace*{5mm}
\noindent
{
In many extensions of the Standard Model, including a broad class
of left-right symmetric and Grand Unified theories, the light neutrino
mass matrix is given by the left-right symmetric seesaw formula
$M_\nu = f v_L - \frac{v^2}{v_R} Y_\nu f^{-1} Y_\nu$,
in which the right-handed neutrino mass matrix and the $SU(2)_L$
triplet couplings are proportional to the same matrix $f$.
We propose a systematic procedure for reconstructing the $2^n$ solutions
(in the $n$-family case) for the matrix $f$ as a function of the Dirac
neutrino couplings $(Y_\nu)_{ij}$ and of the light neutrino mass parameters,
which can be used in both analytical and numerical studies.
We apply this procedure to a particular class of supersymmetric
$SO(10)$ models with two $\bf 10$-dimensional and a pair of
$\bf 126 \oplus \overline{126}$ representations in the Higgs sector,
and study the properties of the corresponding 8 right-handed neutrino
spectra. Then, using the reconstructed right-handed neutrino
and triplet parameters, we study leptogenesis and lepton flavour violation
in these models, and comment on flavour effects in leptogenesis
in the type I limit. We find
that the mixed solutions where both the type I and the type II seesaw
mechanisms give a significant contribution to neutrino
masses provide new opportunities for successful leptogenesis
in $SO(10)$ GUTs.
}
\vspace*{1.0cm}
\date{\today} 


\vspace*{0.2cm}

\vfill\eject

\newpage

\setcounter{page}{1}
\pagestyle{plain}

\section{Introduction}

Experimental data suggest that neutrinos are massive and mix.
The most popular explanation of the smallness of their masses
relies on the (type I) seesaw mechanism~\cite{seesaw},
which finds a natural realization in Grand Unified Theories (GUTs)
based on the $SO(10)$ gauge group.
While the seesaw mechanism cannot be directly tested
since, at least in its GUT version, it involves superheavy states,
it has observable consequences: leptogenesis~\cite{FY86} and,
in supersymmetric theories, flavour~\cite{BM86} and
CP violation~\cite{BHS95} in the lepton sector.
Successful leptogenesis puts several constraints on the seesaw
parameters~\cite{DI02,BDP02,GNRRS03};
in particular, the mass of the lightest right-handed neutrino, $M_1$,
should be larger than ${\cal O} (10^{8} - 10^{9})$ GeV
in the case of a hierarchical mass spectrum.
It is well-known that in $SO(10)$ models where the dominant contribution
to fermion masses comes from 10-dimensional Higgs representations,
the right-handed neutrino mass spectrum reconstructed from the type I
seesaw mass formula is strongly hierarchical (except for special values
of the light neutrino mass parameters~\cite{AFS03}), with $M_1$ lying
below the minimal value for a successful leptogenesis~\cite{ON00}.

However, many extensions of the Standard Model
contain two sources for neutrino masses and leptogenesis: the type I
(associated with the exchange of right-handed neutrinos)~\cite{seesaw}
and type II (associated with the exchange of heavy scalar
$SU(2)_L$ triplets)~\cite{W80,MS81} (see also Ref.~\cite{SW80})
seesaw mechanisms.
In particular, in a broad class of left-right symmetric
and Grand Unified theories, the light neutrino mass matrix
is given by the left-right symmetric seesaw formula
$M_\nu = f v_L - \frac{v^2}{v_R} Y_\nu f^{-1} Y_\nu$, in which
the right-handed neutrino mass matrix and the heavy triplet
couplings are proportional to the same matrix $f$.
Very often in the literature it is assumed that either of these
two mechanisms dominates in the neutrino mass matrix.
But, as discussed below, this corresponds to assuming specific values
of the unknown seesaw parameters. In a  general study one should
encompass the situation where both contributions are sizeable
and can be comparable in magnitude.
In order to implement in an efficient way our experimental
knowledge about neutrino masses and mixings, we need a procedure
to reconstruct the matrix $f$ as a function
of the Dirac neutrino couplings. This problem has been first
addressed in Ref.~\cite{AF05}, where it was found that there are
exactly $2^n$ different solutions in the n-family case, which are
connected two by two by a transformation called ``seesaw duality''.

In this paper, we use a different, more efficient procedure to reconstruct
the $2^n$ solutions, which employs  complex orthogonal transformations
and is appropriate to both numerical and analytic studies.
For three generations of neutrinos, the eight solutions correspond
to the different combinations of the roots of three quadratic equations.
We apply this procedure to a particular class of supersymmetric
$SO(10)$ models with two $\bf 10$-dimensional and a pair of
$\bf 126 \oplus \overline{126}$ representations in the Higgs sector,
and use the results to study leptogenesis and lepton flavour violation
in these models. The spectrum of possibilities to account
for the observed neutrino data is much richer than in the cases
of type I and type II dominance, and the mixed solutions where
both seesaw mechanisms give a significant contribution to neutrino
masses provide new opportunities for successful leptogenesis
in $SO(10)$ GUTs.

The content of the paper is as follows. In Section~\ref{sec:spectrum},
we present the procedure for reconstructing the $2^n$ solutions
for the matrix $f$ from the light neutrino mass parameters,
assuming that the Dirac matrix is known, and we discuss the properties
of the solutions. In section~\ref{sec:SO_10}, we apply the procedure
to supersymmetric $SO(10)$ models with two $\bf 10$-dimensional
and a pair of $\bf 126 \oplus \overline{126}$
representations in the Higgs sector,
and display the corresponding 8 right-handed neutrino spectra
as a function of the $B-L$ breaking scale, for various values
of the free parameters (which include several phases).
In Section~\ref{sec:leptogenesis}, we compute the CP asymmetry
in right-handed neutrino decays for the 8 solutions,
and comment on flavour effects in the type I limit.
In Section~\ref{sec:LFV}, we discuss the predictions
for lepton flavour violating processes.
Finally, in Section~\ref{sec:conclusions}, we give our conclusions
and comment on possible extensions of the present work.
Analytic approximations that can be useful to understand
the results of the reconstruction procedure are given
in Appendix~\ref {app:analytical}.


\section{Reconstruction of the heavy neutrino mass spectrum}
\label{sec:spectrum}

In many extensions of the Standard Model based on a gauge group
embedding the left-right symmetric group
$SU(2)_L \times SU(2)_R \times U(1)_{B-L}$, the neutrino mass matrix
is given by the following seesaw formula~\cite{W80,MS81}:
\beq
  M_\nu\ =\ f_L v_L - \frac{v^2}{v_R} Y^T_\nu f^{-1}_R Y_\nu\ ,
\label{eq:seesaw}
\eeq
where $f_L$ and $f_R$ are symmetric matrices.
The second term in the right-hand side of Eq.~(\ref{eq:seesaw}) is
the usual type I seesaw mass term, where the heavy Majorana
mass matrix $M_R = f_R v_R$  is generated from the vev
of an $SU(2)_R$ triplet $\Delta_R$ with couplings $(f_R)_{ij}$
to right-handed neutrinos, and $M_D = Y_\nu v$ is the Dirac mass matrix
($v = 174$ GeV is the vev of the SM Higgs doublet, to be replaced by
$v_u = v \sin \beta$ in supersymmetric models).
The first term, known as the type II seesaw mass term,
is generated from the exchange of an heavy $SU(2)_L$
triplet $\Delta_L$ with couplings $(f_L)_{ij}$ to lepton doublets.
The induced vev $v_L$ is related to the heavy triplet mass $M_{\Delta_L}$
by $v_L \sim v^2 v_R / M^2_{\Delta_L}$, which naturally explains
its smallness.

In this paper, we consider theories in which the couplings
of the $SU(2)_L$ and $SU(2)_R$ triplets are equal and the
Dirac mass matrix is symmetric, so that Eq.~(\ref{eq:seesaw}) becomes:
\beq
  M_\nu\ =\ f v_L - \frac{v^2}{v_R} Y_\nu f^{-1} Y_\nu\ .
\label{eq:seesaw_LR}
\eeq
These relations arise naturally in $SO(10)$ GUTs in which the right-handed
neutrino masses are generated from a $\bf \overline{126}$ Higgs
representation~\cite{W80} (barring non-symmetric contributions to the Yukawa
couplings, coming e.g. from a $\bf 120$ Higgs representation), as well as
in a broad class of left-right symmetric theories~\cite{MS81}.


\subsection{Reconstruction procedure}
\label{subsec:procedure}

Our starting point is the left-right symmetric seesaw formula~(\ref{eq:seesaw_LR}),
where both $f$ and $Y_\nu$ are complex symmetric matrices.
Our goal is to determine the matrix $f$ for a given pattern of light neutrino
masses and mixings, assuming that the Dirac matrix $Y_\nu$
is known in a basis in which the charged lepton mass matrix is diagonal.
For definiteness we work in the $3$-family case, but the procedure
applies to any number of neutrino families.

If $Y_\nu$ is invertible, Eq. (\ref{eq:seesaw_LR}) can be rewritten in the form:
\beq
  Z\ =\ \alpha X - \beta X^{-1}\ ,
\label{eq:master_equation}
\eeq
with $\alpha \equiv v_L / m_\nu$, $\beta \equiv v^2 / (m_\nu v_R)$ and
\beq
  Z\ \equiv\ \frac{1}{m_\nu}\ Y_\nu^{-1/2} M_\nu (Y_\nu^{-1/2})^T\ , \qquad
  X\ \equiv\ Y_\nu^{-1/2} f (Y_\nu^{-1/2})^T\ ,
\label{eq:def_Z_X}  
\eeq
where $Y_\nu^{1/2}$ is a matrix such that $Y_\nu = Y_\nu^{1/2} (Y_\nu^{1/2})^T$
(in order to deal with dimensionless quantities, we have introduced a mass scale
$m_\nu$ characteristic of the light neutrino mass spectrum). Since $Y_\nu$ is
assumed to be known in the basis of charged lepton mass eigenstates,
the matrix $Z$ is completely determined (up to possible high-energy phases)
by the choice of the light neutrino mass and mixing pattern.

Assuming further that the polynomial equation $\det (Z - z \mathbf{1}) = 0$
has three distinct roots $z_{1,2,3}$, we can diagonalize the complex symmetric
matrix $Z$ with a complex orthogonal matrix\footnote{This is also the case if
$\det (Z - z \mathbf{1}) = 0$ has a multiple root $z_1$, but any non-trivial complex
vector $\vec v$ such that $Z \vec v = z_1 \vec v$ satisfies $\vec v . \vec v \neq 0$.
It should be stressed that, since a complex orthogonal transformation does {\it not}
preserve the norm of states, $|\vec v|^2 \equiv \vec v^\star . \vec v$,
Eq. (\ref{eq:diag_Z}) is not a diagonalization in the physical sense.}:
\beq
 Z\ =\ O_Z \mbox{Diag}\, (z_1, z_2, z_3) O^T_Z\ ,  \qquad  O_Z O^T_Z\ =\ \mathbf{1}\ ,
\label{eq:diag_Z} 
\eeq
where the $z_i$ are complex numbers.
Then Eq. (\ref{eq:master_equation}) can be solved for $X$ in a straightforward
manner, by noting that $X$ is diagonalized by the same complex orthogonal
matrix as $Z$:
\beq
 X\ =\ O_Z \mbox{Diag}\, (x_1, x_2, x_3) O^T_Z\ ,
\label{eq:diag_X}
\eeq
with the $x_i$ being the solutions of the quadratic equation
$z_i = \alpha x_i - \beta x^{-1}_i$. 
For a given choice of ($x_1$, $x_2$, $x_3$), the matrix $f$ is given by:
\beq
  f\ =\ Y_\nu^{1/2} X (Y_\nu^{1/2})^T\ =\
  Y_\nu^{1/2}\, O_Z\, \mbox{Diag}\, (x_1, x_2, x_3)\, O^T_Z\, (Y_\nu^{1/2})^T\ ,
\label{eq:f}
\eeq
and the right-handed neutrino masses
$M_i = f_i v_R$ are obtained by diagonalizing $f$ with a unitary matrix:
\beq
  f\ =\ U_f \hat f U^T_f\ , \qquad  \hat f\ =\ \mbox{Diag}\, (f_1, f_2, f_3)\ ,
  \qquad  U_f U^\dagger _f = \mathbf{1}\ ,
\label{eq:diag_f}
\eeq
where the $f_i$ are chosen to be real and positive.
The matrix $U_f$ relates the original basis for right-handed neutrinos,
in which $Y_\nu$ is symmetric, to their mass eigenstate basis. It can be
used to express the Dirac couplings in terms of charged lepton
and right-handed neutrino mass eigenstates, as
$Y \equiv U^\dagger_f Y_\nu$.  

Since there are two possible choices for each $x_i$, we have 8 different
solutions for the matrix $f$ ($2^n$ in the $n$-generation case),
a property already found in Ref.~\cite{AF05}.
The advantage of the above procedure
over the one presented in Ref.~\cite{AF05} is that it is more systematic and
allows for an easier reconstruction of the right-handed neutrino masses
and mixings, both analytically and numerically. Furthermore, the connection
between the different solutions is more transparent.


\subsection{Properties of the 8 solutions}
\label{subsec:properties}

In order to label the 8 different solutions for $f$, we denote the two solutions
of the equation $z_i = \alpha x_i - \beta x^{-1}_i$ by $x^+_i$ and $x^-_i$, with
\beq
  x^\pm_i\ \equiv\ \frac{z_i \pm \mbox{sign} (\mbox{Re} (z_i))
    \sqrt{z^2_i + 4 \alpha \beta}}{2 \alpha}\ .
\label{eq:x_pm_i}
\eeq
With the above definition, one has, in the limit where $4 \alpha \beta \ll |z_i|^2\, $:
\beq
  x^+_i\ \simeq\ \frac{z_i}{\alpha}\ ,  \qquad  x^-_i\ \simeq\ - \frac{\beta}{z_i}\ ,
\label{eq:x_pm_limit}
\eeq
while in the limit $|z_i|^2 \ll 4 \alpha \beta$:
\beq
  x^\pm_i\ \simeq\ \pm\, \mbox{sign} (\mbox{Re} (z_i)) \sqrt{\beta / \alpha}\ .
\eeq
We label the 8 solutions for $f$ in the following way: $(+,+,+)$ refers
to the solution $(x^+_1, x^+_2, x^+_3)$, $(+,+,-)$ to the solution
$(x^+_1, x^+_2, x^-_3)$, and so on.
We will sometimes refer to $x^-_i$ as the ``type I branch'' and to $x^+_i$
as the ``type II branch'', for reasons that will become clear below.

For some particular solutions, Eq. (\ref{eq:x_pm_limit}) has an immediate
interpretation in terms of dominance of the type I or type II contribution
to light neutrino masses. More precisely, for values of $\alpha \beta$
such that $4 \alpha \beta \ll |z_1|^2$, the solutions $(+,+,+)$ and $(-,-,-)$
practically coincide with the ``pure'' type II and type I cases, respectively.
Indeed, plugging the approximate relations $X^{(+,+,+)} \simeq Z / \alpha$
and $X^{(-,-,-)} \simeq -\beta Z^{-1}$ into Eqs. (\ref{eq:def_Z_X}), one obtains:
\bea
  f^{(+,+,+)} & \stackrel{4 \alpha \beta \ll |z_1|^2}{\longrightarrow} &
  \frac{M_\nu}{v_L}  \qquad \qquad \qquad \ \  \mbox{(type II limit)}\ ,  \\ 
  f^{(-,-,-)} & \stackrel{4 \alpha \beta \ll |z_1|^2}{\longrightarrow} &
  -\, \frac{v^2}{v_R}\, Y_\nu M^{-1}_\nu Y_\nu  \qquad  \mbox{(type I limit)}\ .
\eea
The remaining 6 solutions correspond to mixed cases where, even
in the $4 \alpha \beta \ll |z_1|^2$ limit, the light neutrino mass matrix
receives significant contributions from both types
of seesaw mechanisms. Depending on the Dirac couplings as well as
on the light neutrino mass and mixing pattern, one may have solutions
where e.g. the type I contribution dominates in some entries of the
light neutrino mass matrix, while both seesaw mechanisms contribute
to the other entries\footnote{In fact, analogously to the type I case in which the
complex orthogonal matrix $R$ introduced by Casas and Ibarra~\cite{CI01}
in order to parametrize the seesaw mechanism can be interpreted as a
``dominance matrix''~\cite{LMS02}, one may define a dominance matrix
$Q \equiv \sqrt{m_\nu}\, (Z^{1/2})^T (Y^{1/2}_\nu)^T (M^{-1/2}_\nu)^T$
such that $m_i = \sum_k Q^2_{ki} m_i$, where $Q^2_{ki} m_i$ is the contribution
of $z_k$ to $m_i$. If e.g. the contribution of $z_1$ dominates in $m_3$
(i.e. $|Q_{13}|^2 \gg |Q_{23}|^2, |Q_{33}|^2$),
then one can say that either the type I or the type II contribution dominates
in $m_3$, depending on whether $x_1 = x^-_1$ or $x_1 = x^+_1$.}.

There is another range of values for $\alpha \beta$ in which $f$ reaches
a remarkable limit, namely $|z_3|^2 \ll 4 \alpha \beta$. In this region, one has
$x^\pm_i \simeq \pm \mbox{sign} (\mbox{Re} (z_i)) \sqrt{\beta / \alpha}$
for all $i$, which indicates a strong cancellation between the type I
and type II contributions to the light neutrino mass matrix. 
This can easily be seen for the two solutions labelled by
$(\pm \epsilon_1, \pm \epsilon_2, \pm \epsilon_3)$,
$\epsilon_i \equiv \mbox{sign} (\mbox{Re} (z_i))$, in which one has
$X \simeq \pm \sqrt{\beta / \alpha}\, {\bf 1}$.
Using Eqs.~(\ref{eq:def_Z_X}), this leads to:
\beq
  f^{(\pm \epsilon_1, \pm \epsilon_2, \pm \epsilon_3)}\
  \stackrel{4 \alpha \beta \gg |z_3|^2}{\longrightarrow}\
  \pm \sqrt{\beta / \alpha}\, Y_\nu\ ,
\label{eq:f_cancel}
\eeq
which shows that the type I and type II contributions
approximately cancel in $M_\nu$.
For the other 6 solutions, Eq.~(\ref{eq:f_cancel}) does not hold
but one still has $f_i \simeq \sqrt{\beta / \alpha}\, y_i$ (where the $y_i$
are the eigenvalues of $Y_\nu$) in the $|z_3|^2 \ll 4 \alpha \beta$ regime,
provided that the Dirac matrix has a hierarchical structure. Moreover,
for the type of hierarchy considered in Section~\ref{sec:SO_10},
one can show that $f \simeq \sqrt{\beta / \alpha}\,
U^T_\nu \mbox{Diag}\, (s_1 y_1, s_2 y_2, s_3 y_3) U_\nu$
in the $|z_3|^2 \ll 4 \alpha \beta$ regime, where
$Y_\nu = U^T_\nu \mbox{Diag}\, (y_1, y_2, y_3) U_\nu$ and
$s_i = \pm\, \mbox{sign} (\mbox{Re} (z_i))$ for $x_i = x^\pm_i$
(see Appendix~\ref{app:analytical}).

Finally, in the intermediate region of values for $\alpha \beta$,
$|z_1|^2 < 4 \alpha \beta < |z_3|^2$,
both the type I and the type II seesaw mechanism give
significant contributions to the light neutrino mass matrix.
Already for $|z_1|^2 \ll 4 \alpha \beta$, cancellations between
the right-handed neutrino and triplet contributions to $M_\nu$
start to occur.

The 8 different solutions for $f$ are connected to each other by the
three transformations:
\beq
  x_i\ \rightarrow\ \tilde x_i \equiv z_i / \alpha - x_i\ ,
\label{eq:duality}
\eeq
which act as $x^+_i \leftrightarrow x^-_i$.
This generalizes the ``seesaw duality'' of Ref.~\cite{AF05}, defined as
$f \rightarrow \tilde f \equiv M_\nu / v_L - f$, which amounts
to interchange the type I and type II branches for all three $x_i$
simultaneously, thus dividing the 8 solutions into 4 ``dual pairs''.
The transformations (\ref{eq:duality}), more generally, allow to generate
all 8 solutions from a single one. 
In group-theoretical terms, these transformations define 
an abelian group $Z_{2}\otimes Z_{2}\otimes Z_{2}$ of 8 elements,
one of which is the ``seesaw duality'' of  Ref.~\cite{AF05}.


\section{A case study: right-handed neutrinos in $SO(10)$ models}
\label{sec:SO_10}

The procedure described in Subsection~\ref{subsec:procedure}
can be used to determine the a priori unknown $f_{ij}$ couplings
in theories which predict the Dirac matrix $Y_\nu$, taking low-energy
neutrino data as an input. In the following, we apply it to reconstruct
the right-handed neutrino mass spectrum in  a class of supersymmetric
$SO(10)$ models with two $\bf 10$-dimensional and a pair of
$\bf  126 \oplus \overline{126}$ representations in the Higgs sector.


\subsection{Input parameters}
\label{subsec:models}

In supersymmetric $SO(10)$ models with two $\bf 10$-dimensional
and a pair of $\bf  126 \oplus \overline{126}$ representations (but no
$\bf 120$-dimensional representation) in the Higgs
sector, the most general Yukawa couplings read:
\beq
  Y^{(1)}_{ij}\, {\bf 16_i} {\bf 16_j} {\bf 10_1} + Y^{(2)}_{ij}\, {\bf 16_i} {\bf 16_j} {\bf 10_2}
  + f_{ij}\, {\bf 16_i} {\bf 16_j} {\bf \overline{126}}\ ,
\label{eq:Yukawa_SO10}
\eeq
where $Y^{(1)}$, $Y^{(2)}$ and $f$ are complex symmetric matrices.
Assuming that the $SU(2)_L$ doublet components of the $\bf \overline{126}$
do not acquire a vev, Eq.~(\ref{eq:Yukawa_SO10}) leads to the following
mass relations for the charged fermions, valid at the GUT scale:
\beq
  M_u\ =\ M_D\ , \qquad M_d\ =\ M_e\ .
\label{eq:M_10}
\eeq
It is well-known that the second relation is in conflict with experimental data
and needs to be corrected. In general, the corrections (coming e.g. from the
$SU(2)_L$ doublet components in the $\bf \overline{126}$~\cite{MS80}, or from
non-renormalizable operators~\cite{ADHRS94}) also affect the first relation.
Although these corrections will change numerically the solutions for $f$,
we do not expect them to alter the qualitative features of our results.
As a case study, we  assume Eq.~(\ref{eq:M_10}) to hold in the following.

The inputs in the procedure for reconstructing the 8 solutions
for $f$ are the matrices $Y_\nu$ and $M_\nu$
at the seesaw scale\footnote{Actually Eq.(\ref{eq:seesaw_LR})
involves the decoupling of four states at scales that can differ
by several orders of magnitude; the associated radiative corrections
are neglected here.}. 
The ``boundary condition'' for $Y_\nu$, Eq.~(\ref{eq:M_10}),
is defined at the GUT scale, where it is convenient to work in the basis
for the $\bf 16$ matter representations in which $M_e$ (hence $M_d$)
is diagonal with real positive entries. In this basis, the Dirac matrix reads:
\beq
  Y_\nu\ =\ U^T_q \hat Y_u U_q\ , \qquad U_q\ =\ P_u V_{CKM} P_d\ ,
  \qquad \hat Y_u\ =\ \mbox{Diag}\, (y_u, y_c, y_t)\ ,
\label{eq:Y_nu}
\eeq
where $V_{CKM}$ is the CKM matrix and $y_{u, c, t}$ are the up quark
Yukawa couplings, all renormalized at the GUT scale.
The presence of two diagonal matrices of phases $P_u$ and $P_d$
in Eq.~(\ref{eq:Y_nu}) is due to the fact that the $SO(10)$ symmetry prevents
independent rephasing of right-handed and left-handed quark fields.
Since $Y_\nu$ is only weakly renormalized between $M_{GUT}$ and
the seesaw scale, the effet of the running being smaller than the
uncertainty on the quark parameters at $M_{GUT}$, we can neglect it
and assume Eq.~(\ref{eq:Y_nu}) to hold at the seesaw scale,
in the basis of charged lepton mass eigenstates.
In the same basis, the light neutrino mass matrix generated
from the seesaw mechanism reads:
\beq
  M_\nu\ =\ U^\star_l \hat M_\nu U^\dagger_l\ , \qquad U_l\ =\ P_e U_{PMNS} P_\nu\ ,
  \qquad \hat M_\nu\ =\ \mbox{Diag}\, (m_1, m_2, m_3)\ ,
\label{eq:M_nu}
\eeq
where $U_{PMNS}$ is the PMNS matrix and $m_{1,2,3}$ are the light
neutrino masses, all renormalized at the seesaw scale.
The two relative phases in $P_\nu$ are the physical CP-violating phases
associated with the Majorana nature of the light neutrinos,
while the three phases contained in $P_e$, analogous to the five independent
phases contained in $P_u$ and $P_d$, are pure high-energy phases.
Having specified Eqs.~(\ref{eq:Y_nu}) and~(\ref{eq:M_nu}), one can
apply the procedure presented in Subsection~\ref{subsec:procedure}
and reconstruct the 8 different matrices $f$ corresponding to a given
light neutrino mass and mixing pattern as a function of $\alpha$, $\beta$
and of the high-energy phases contained in $P_u$, $P_d$ and $P_e$.

The associated right-handed neutrino mass and mixing patterns strongly
depend on the values of $\alpha$ and $\beta$ (or equivalently $\beta / \alpha$
and $v_R$), which in turn depend on the details of the model.
The simplest way to realize the type II seesaw mechanism in the class
of $SO(10)$ models considered is to introduce a $\bf 54$ representation
in the Higgs sector in addition to the $\bf \overline{126} \oplus 126$ pair.
This is easily seen in a left-right symmetric
language: the $\bf \overline{126}$ contains a right-handed triplet
$\Delta^c$ with quantum numbers ${\bf (1,3,1)_{-2}}$ under
$SU(2)_L \times SU(2)_R \times SU(3)_C \times U(1)_{B-L}$, whose vev
$v_R$ is responsible for the breaking of $B-L$,
as well as a left-handed triplet $\Delta = {\bf (3,1,1)_{+2}}$; the $\bf 54$
contains a bitriplet $\tilde \Delta = {\bf (3,3,1)_0}$; and each $\bf 10$
contains a bidoublet $\Phi = {\bf (2,2,1)_0}$.
The superpotential terms relevant for the type II contribution
to neutrino masses include:
\beq
  W\ =\ \frac{1}{2}\, f_{ij}\, L_i L_j \Delta
  + \kappa\, \Phi \Phi \tilde \Delta
  + \lambda\, \Delta \Delta^c \tilde \Delta +\ \ldots\ ,
\label{eq:W_typeII}
\eeq
where the first term comes from the ${\bf 16_i} {\bf 16_j} {\bf \overline{126}}$
couplings, and the second and third terms come from the $\bf 10\, 10\, 54$
and $\bf 54\, \overline{126}\, \overline{126}$ couplings, respectively.
The presence of these terms induces
a vev $v_L \sim \kappa \lambda v^2_u v_R / M^2_{\Delta_L}$
for the $SU(2)_L$ triplet, yielding a ratio $\beta / \alpha =  v^2_u / (v_L v_R)
\sim M^2_{\Delta_L} / (\kappa \lambda v^2_R)$.
Depending on the other superpotential couplings, the triplet mass
$M_{\Delta_L}$ may be larger or smaller than $v_R$ (for $v_R \ll M_{GUT}$,
a tuning of the superpotential parameters might be necessary
to achieve\footnote{Due to the $\bf 54\, \overline{126}\, \overline{126}$
interaction term, the $SU(2)_L$ triplet states in the $\bf 54$ and
$\bf \overline{126}$ representations mix below $v_R$. By suitably
tuning the values of the superpotential couplings, one can decrease
the mass of the lightest eigenvalue of the triplet mass matrix~\cite{GMN04}.}
$M_{\Delta_L} < v_R$), hence $\beta / \alpha$ can be larger or smaller than $1$.
As for $v_R$, its value is related to the breaking scheme of the GUT
symmetry and is also model dependent. 
In principle the $B-L$ symmetry could be broken anywhere
between the Planck scale and the weak scale;
however the requirement of gauge coupling unification generically
disfavours the breaking of $B-L$ at lower scales.
Detailed studies (see e.g. Ref.~\cite{ABMRS01}) have shown that
the $B-L$ symmetry can be broken a few orders of magnitude
below the GUT scale consistently with unification. In the following,
we allow $v_R$ to vary in the range $(10^{12} - 10^{17})$ GeV.


\subsection{Right-handed neutrino spectra}
\label{subsec:spectra}

In this subsection, we display the right-handed neutrino spectra obtained
from the reconstruction of the couplings $f_{ij}$ in the class of $SO(10)$
models specified above. For definiteness,
we consider the case of a hierarchical light neutrino mass spectrum with
$m_1 = 10^{-3}$ eV, and we take the best fit values of Ref.~\cite{fits}
for the oscillation parameters (the parametrization of the PMNS matrix
is the one adopted in the Review of Particle Properties~\cite{PDG04}):
\bea
  & \Delta m^2_{32} \equiv m^2_3 - m^2_2 = 2.4 \times 10^{-3}\, \mbox{eV}^2 , \qquad
  \Delta m^2_{21} \equiv m^2_2 - m^2_1 = 7.92 \times 10^{-5}\, \mbox{eV}^2 , &
  \label{eq:masses_fit}  \\
  & \sin^2 \theta_{23} = 0.44\ , \qquad \sin^2 \theta_{12} = 0.314\ , \qquad
  \sin^2 \theta_{13} = 0.009\ . &
\label{eq:angles_fit}
\eea
The PMNS phase $\delta$ and the two relative Majorana phases
contained in $P_\nu$ are treated as free parameters in our study,
like the high-energy phases contained in $P_u$, $P_d$ and $P_e$.
The renormalization group running between low energy
and the seesaw scale has little impact on the neutrino parameters
in the case of a hierarchical spectrum, apart from an overall scaling
of the light neutrino masses~\cite{RGEnu} which we roughly take into account
by multiplying their low-energy values by a factor $1.2$~\cite{AKLR03}.

In the quark sector, we take into account the renormalization group running
by setting $A (M_{GUT}) = 0.7$ in the Wolfenstein parametrization of the CKM
matrix, and $\hat Y_u (M_{GUT}) =
\mbox{Diag}\, (6 \times 10^{-6}, 2.5 \times 10^{-3}, 1) \times y_t (M_{GUT})$,
with $y_t (M_{GUT}) = 0.7$. In our estimate of the GUT-scale values
for the up quark Yukawa couplings, we have taken the central values
for the first and second generation quark masses given in the
Review of Particle Properties~\cite{PDG04}.
Furthermore, we take $\lambda = 0.22$, $\rho = 0.2$ and $\eta = 0.35$,
in agreement with fits of the unitary triangle~\cite{ckmfitter}.

Before presenting the 8 right-handed neutrino spectra corresponding
to these inputs, let us mention the restrictions that apply to the reconstructed
couplings $f_{ij}$. A first restriction comes from the requirement
of perturbativity, i.e. the $f_{ij}$'s should remain in the perturbative regime
up to the scale at which the unified gauge coupling $g_{10}$ blows up,
$\Lambda_{10} \approx 2 \times 10^{17}$ GeV. As discussed in
Appendix~\ref{app:perturbativity}, we can safely take $f_3 < 1$
as a perturbativity constraint at the seesaw scale where the $f_{ij}$'s
are determined.
One can impose a second restriction on the couplings $f_{ij}$ by requiring that there
be no unnatural cancellations between the type I and the type II contributions to
neutrino masses. In practice, we shall define the fine-tuned region by:
\beq
  f_{33}\, v_L\ >\ F (M_\nu)_{33}\ , 
\eeq
where $F$ measures the level of fine-tuning in the $(3, 3)$ entry of the light neutrino
mass matrix: $F=10$ corrresponds to a $10 \%$ fine-tuning, and so on.
Such cancellations might be the consequence of a symmetry ensuring
a proportionality relation between $f$ and $Y_\nu$,
and are not necessarily unnatural. Nevertheless a high degree of fine-tuning
would be unstable against radiative corrections, since this symmetry must
be broken.

We are now ready to display the 8 solutions for the couplings $f_{ij}$
(or, more precisely, the right-handed neutrino masses $M_i = f_i v_R$ and the entries
of the unitary matrix $U_f$ that diagonalizes $f$) as a function of $v_R$,
for a given value of $\beta / \alpha$ and of the phases $\delta$, $\Phi^u_i$, $\Phi^d_i$,
$\Phi^\nu_i$ and $\Phi^e_i$ ($i=1,2,3$). We take as a reference point the case in which
the light neutrino mass spectrum is hierarchical with $m_1 = 10^{-3}$ eV,
$\beta = \alpha$ (i.e. $v_L = v^2_u / v_R$), and the Yukawa matrices do not contain
any CP-violating phase beyond the CKM phase.
We allow $v_R$ to vary between $10^{12}$ GeV and $10^{17}$ GeV but,
as we shall see,
the perturbativity constraint $f_3 < 1$ generally restricts this range.

\begin{figure}
\begin{center}
\includegraphics*[height=4.5cm]{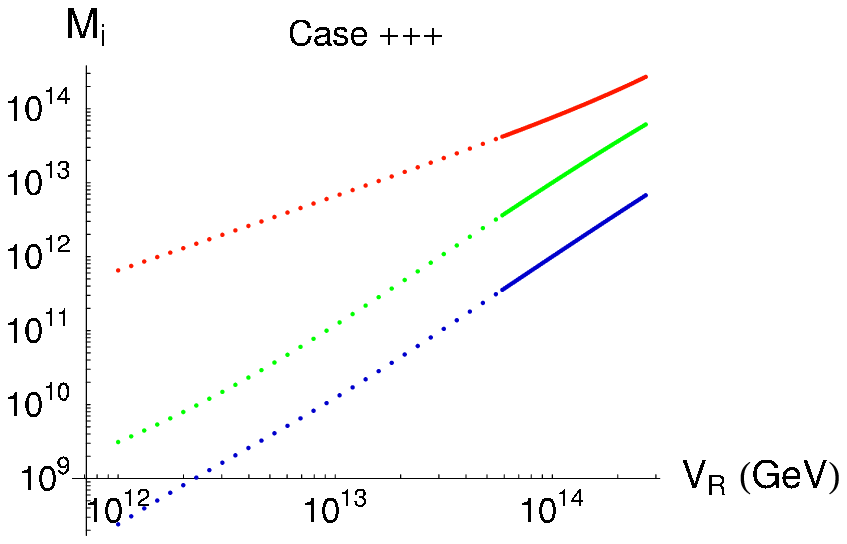}
\hskip 1cm
\includegraphics*[height=4.5cm]{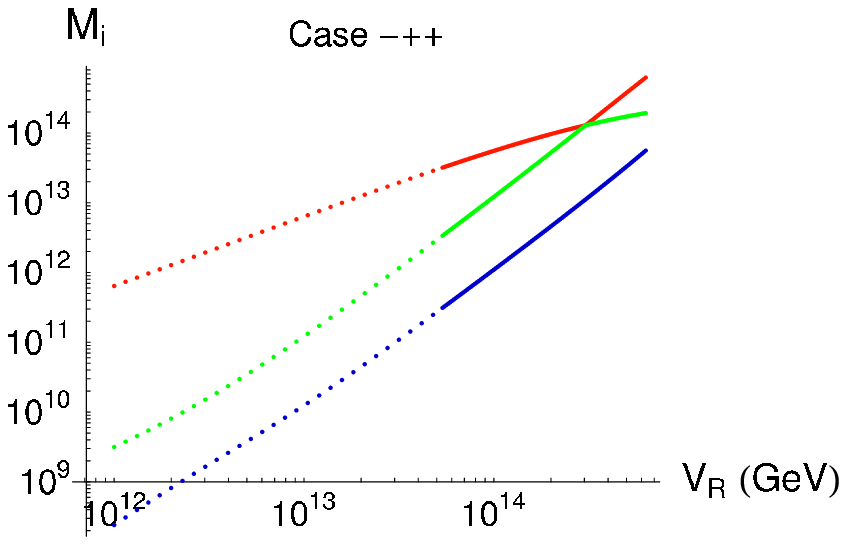}
\vskip .5cm
\includegraphics*[height=4.5cm]{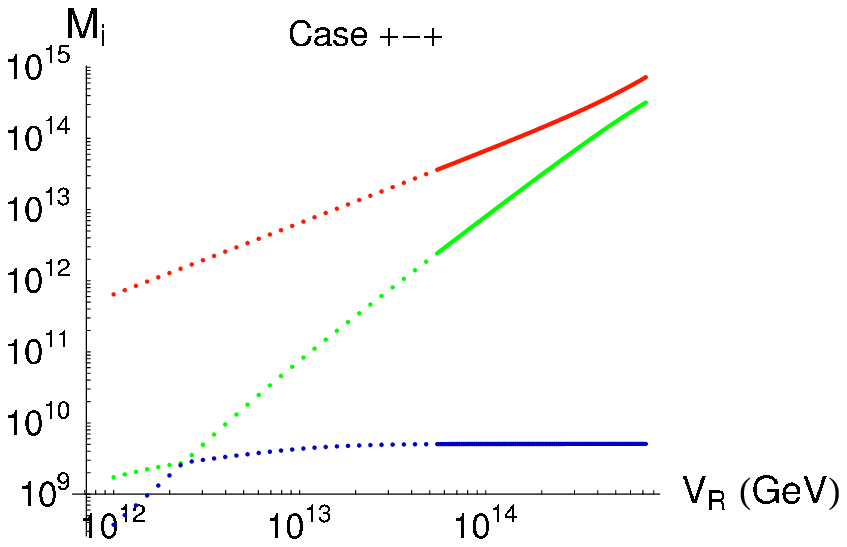}
\hskip 1cm
\includegraphics*[height=4.5cm]{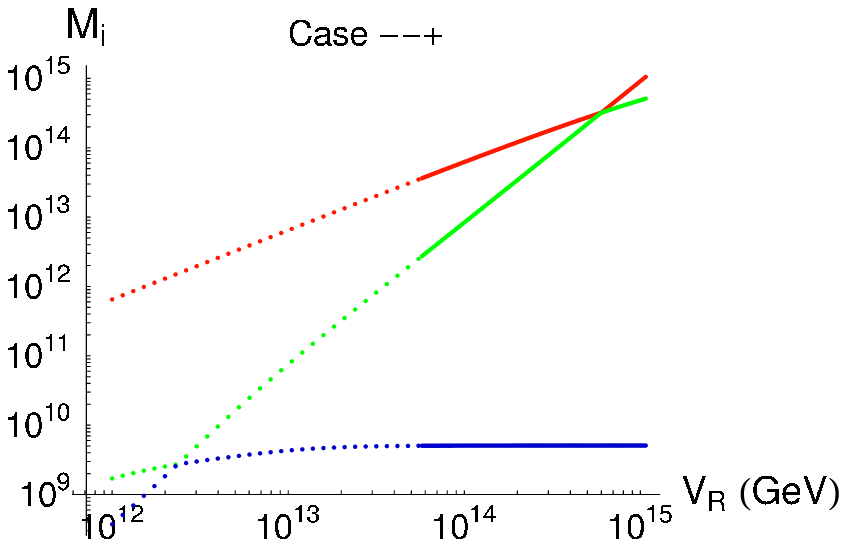}
\vskip .5cm
\includegraphics*[height=4.5cm]{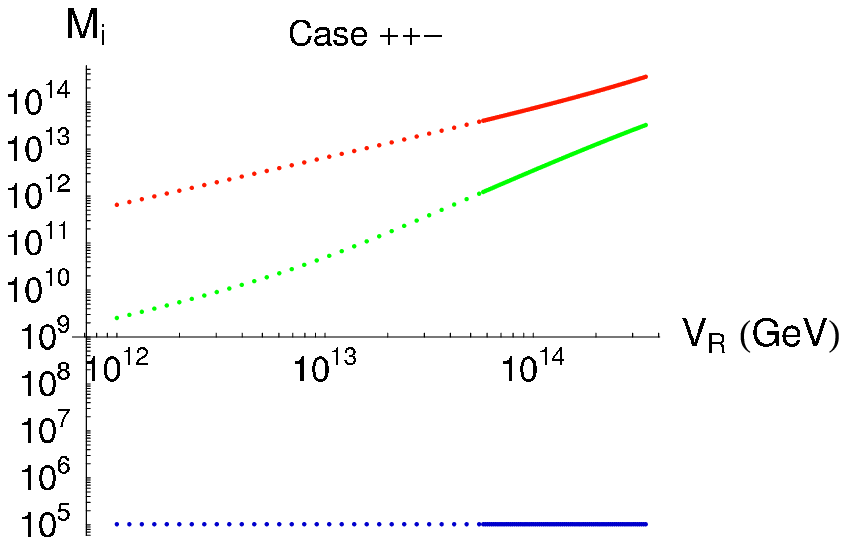}
\hskip 1cm
\includegraphics*[height=4.5cm]{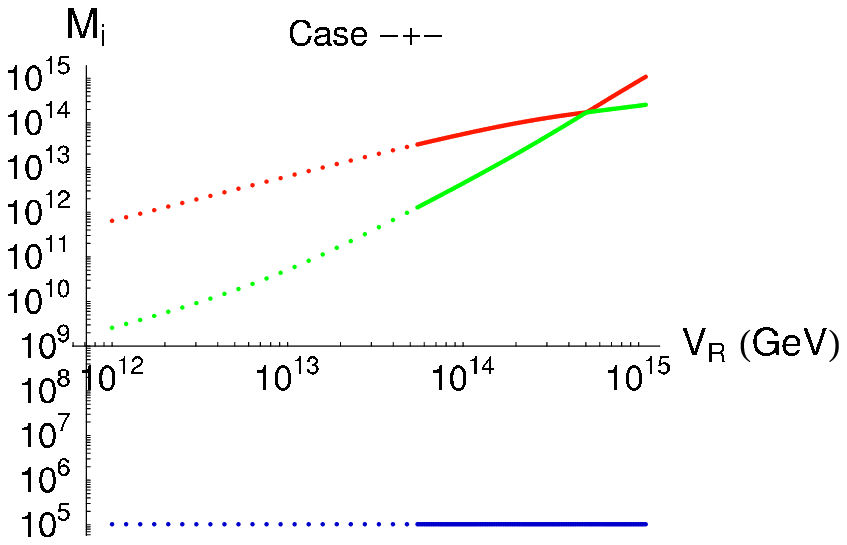}
\vskip .5cm
\includegraphics*[height=4.5cm]{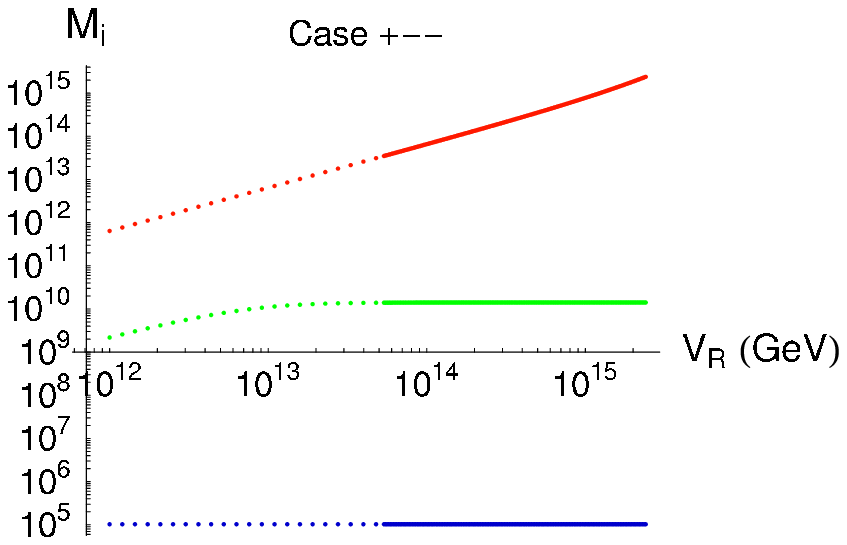}
\hskip 1cm
\includegraphics*[height=4.5cm]{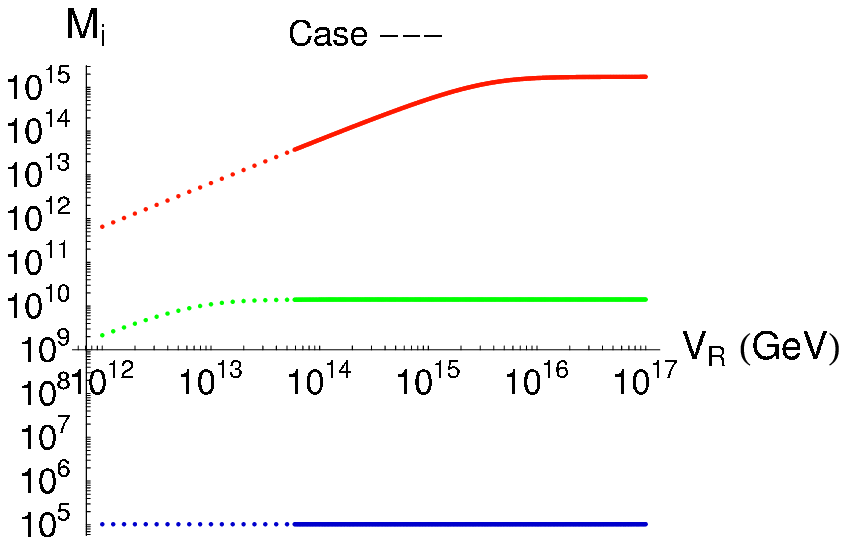}
\vskip .3cm
\caption{Right-handed neutrino masses as a function of $v_R$ for each
of the 8 solutions $(+,+,+)$ to $(-,-,-)$ in the reference case of a hierarchical
light neutrino mass spectrum with $m_1 = 10^{-3}$ eV, $\beta = \alpha$
and no CP violation beyond the CKM phase
($\delta = \Phi^u_i = \Phi^d_i = \Phi^\nu_i = \Phi^e_i = 0$). The range of
variation of $v_R$ is restricted by the requirement that $f_3 \leq 1$.
Dotted lines indicate a fine-tuning greater than $10 \%$ in the $(3,3)$
entry of the light neutrino mass matrix.}
\label{fig:Mi_ref}
\end{center}
\end{figure}

\begin{figure}
\begin{center}
\includegraphics*[height=4.5cm]{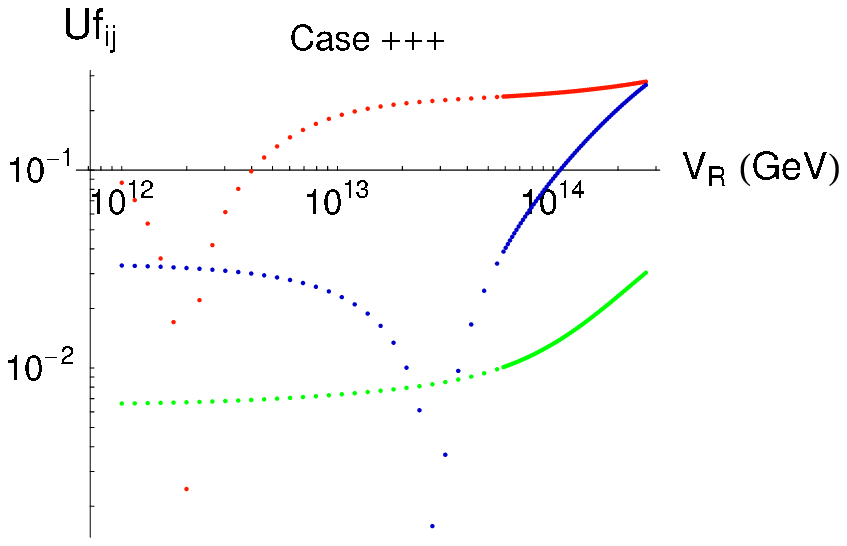}
\hskip 1cm
\includegraphics*[height=4.5cm]{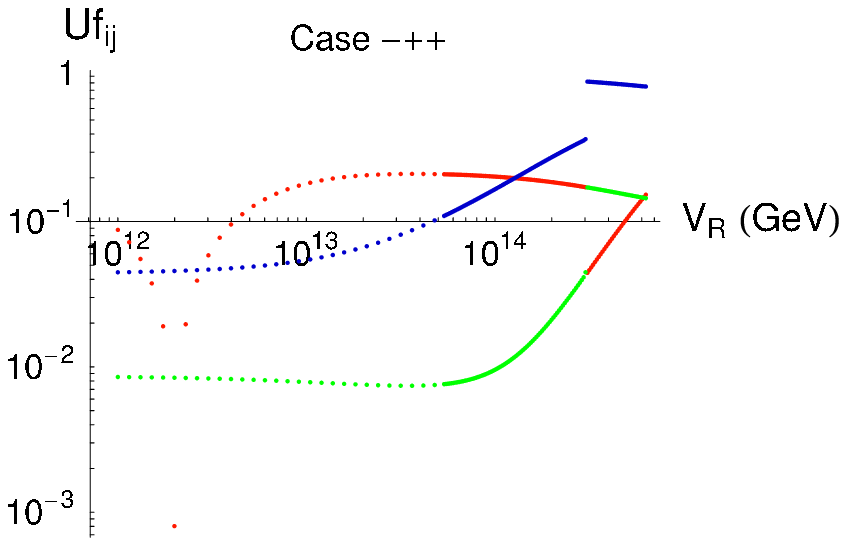}
\vskip .5cm
\includegraphics*[height=4.5cm]{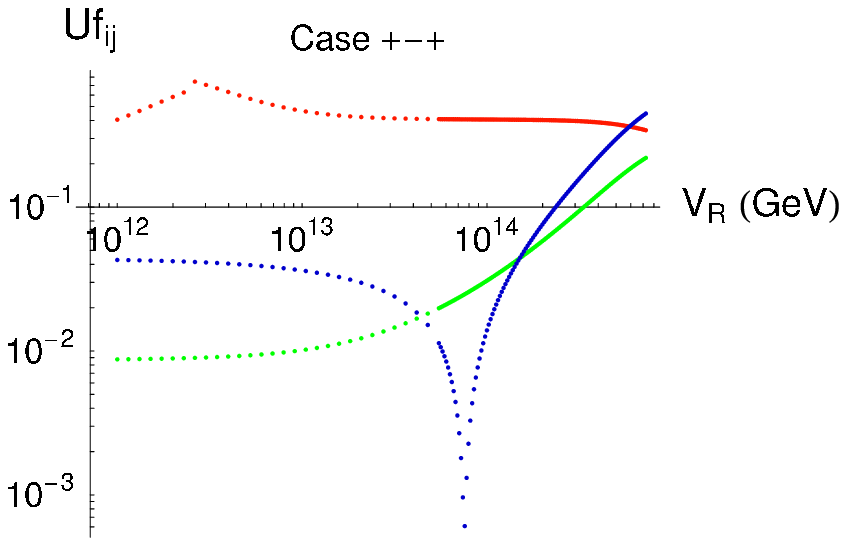}
\hskip 1cm
\includegraphics*[height=4.5cm]{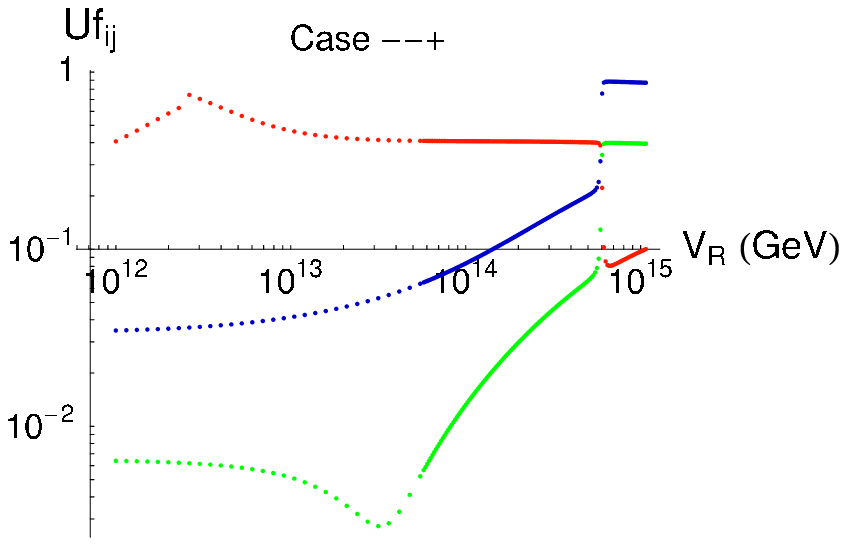}
\vskip .5cm
\includegraphics*[height=4.5cm]{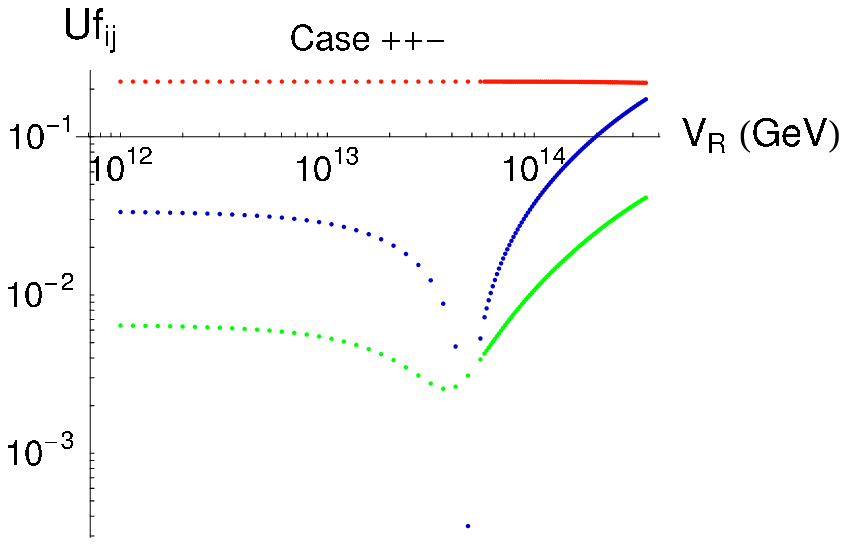}
\hskip 1cm
\includegraphics*[height=4.5cm]{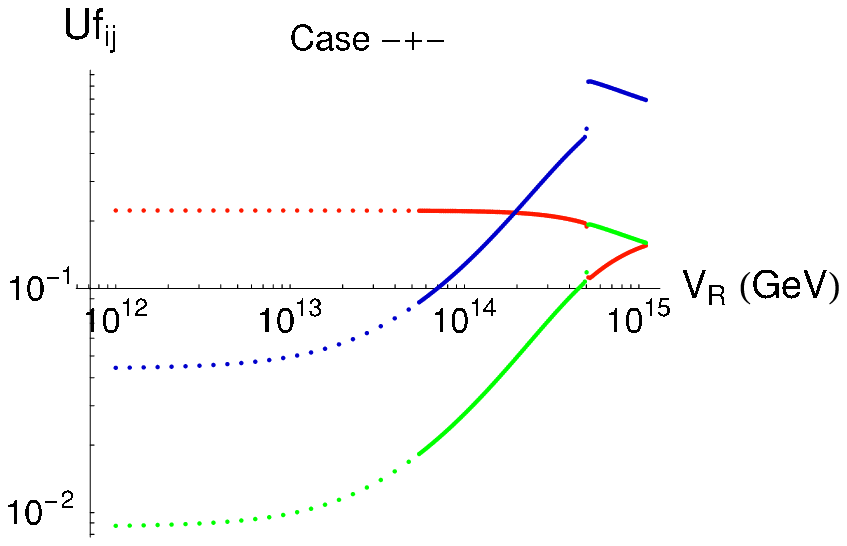}
\vskip .5cm
\includegraphics*[height=4.5cm]{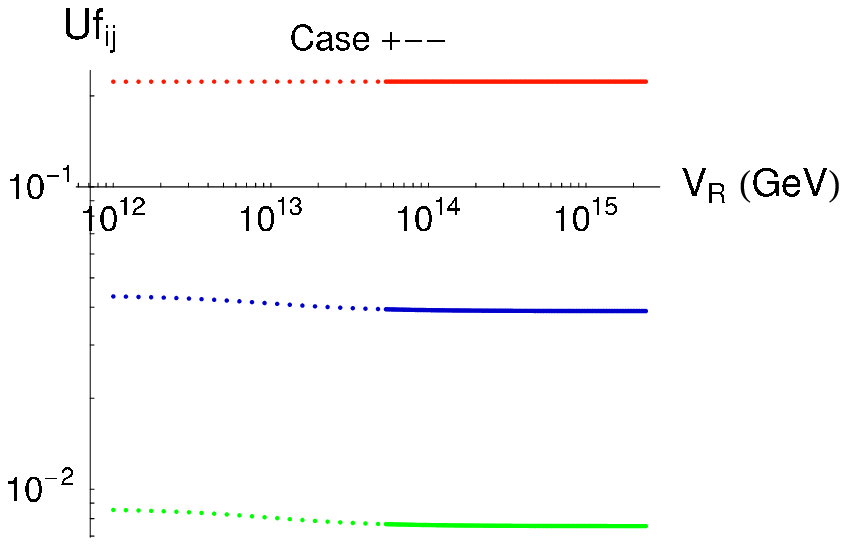}
\hskip 1cm
\includegraphics*[height=4.5cm]{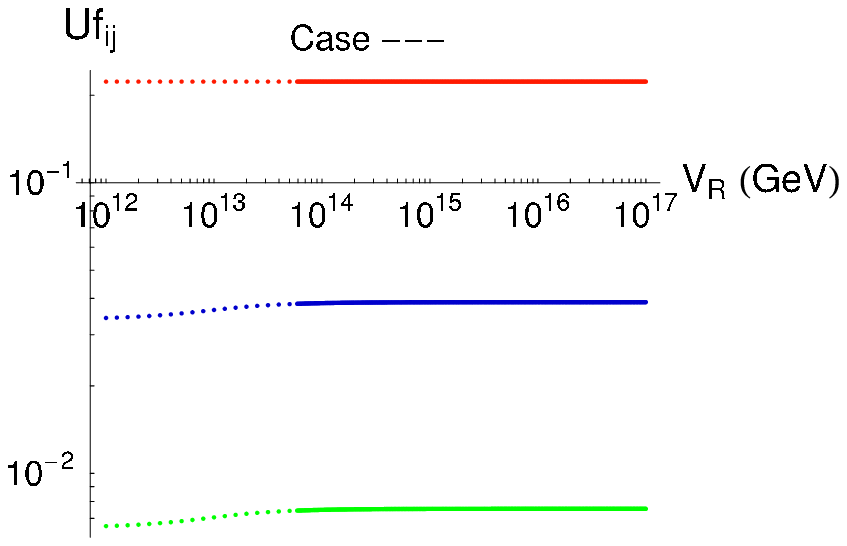}
\vskip .3cm
\caption{Right-handed neutrino mixing angles as a function of $v_R$ for each
of the 8 solutions $(+,+,+)$ to $(-,-,-)$ in the reference case of a hierarchical
light neutrino mass spectrum with $m_1 = 10^{-3}$ eV, $\beta = \alpha$
and no CP violation beyond the CKM phase
($\delta = \Phi^u_i = \Phi^d_i = \Phi^\nu_i = \Phi^e_i = 0$).
The red [dark grey] curve corresponds to $|(U_f)_{12}|$, the green [light grey] curve
to $|(U_f)_{13}|$, and the blue [black] curve to $|(U_f)_{23}|$.}
\label{fig:Uij_ref}
\end{center}
\end{figure}

Figs.~\ref{fig:Mi_ref} and \ref{fig:Uij_ref} show the right-handed neutrino masses
$M_{1,2,3}$ and their mixing angles $|(U_f)_{12}|$, $|(U_f)_{13}|$
and $|(U_f)_{23}|$ as a function of $v_R$, for each of the 8 solutions
$(+,+,+)$ to $(-,-,-)$, in the reference case $m_1 = 10^{-3}$ eV, $\beta = \alpha$
and $\delta = \Phi^u_i = \Phi^d_i = \Phi^\nu_i = \Phi^e_i = 0$.
Due to the interplay between the type I and type II contributions, the observed
light neutrino masses and mixing angles are compatible with a large variety
of right-handed neutrino mass spectra.
As discussed in Subsection \ref{subsec:properties},
one recovers the type I spectrum, characterized by the approximate hierarchy
$M_1 : M_2 : M_3 \sim m^2_u : m^2_c : m^2_t$, in the large $v_R$ region
of solution $(-,-,-)$. Each eigenvalue $M_i$ reaches its type I value
when the condition $4 \alpha \beta \ll |z_{4-i}|^2$ is satisfied, i. e.
when $v_R \gg 2 v^2_u / |z_{4-i}|$. This explains why $M_1$ is constant over
the considered range of values for $v_R$, while $M_2$ (resp. $M_3$)
reaches a plateau only above $v_R \sim 10^{13}$ GeV (resp.
$v_R \sim 5 \times 10^{15}$ GeV). Symmetrically, the type II limit,
characterized by the mild hierarchy $M_1 : M_2 : M_3 \propto m_1 : m_2 : m_3$,
corresponds to the large $v_R$ region of solution $(+,+,+)$. Finally, in the
small $v_R$ region ($v_R \lesssim 10^{10}$ GeV)
where there is a very strong cancellation between the type I
and type II contributions, one finds an intermediate hierarchy
$M_1 : M_2 : M_3 \propto m_u : m_c : m_t$
(see Eq.~(\ref{eq:f_cancel}) and discussion below).
Although this region is not shown in the figures, one can already see
that $M_2 : M_3 \sim m_c : m_t$ for $v_R = 10^{12}$ GeV.
These features of the right-handed neutrino mass spectra are expected
to hold in other models where the Dirac mass matrix has a strong
hierarchical structure, but is not necessarily related to the up quark
mass matrix like in $SO(10)$ models.

The differences between the 8 different spectra plotted in Fig.~\ref{fig:Mi_ref} 
can be understood by noting that each $M_i$ can be associated (in the sense
explained in Appendix~\ref{app:analytical}) with one of the eigenvalues of the matrix
$X$, say $x_j$.
If $x_j = x^-_j$ (``type I branch''), the corresponding right-handed neutrino
mass $M_i$ first grows linearily with $v_R$, then reaches a plateau for
$v_R \gg 2 \sqrt{\alpha / \beta}\, v^2_u / |z_j|$, corresponding to $x_j \simeq - \beta / z_j$.
If $x_j =x^+_j$ (``type II branch''), $M_i$ first grows linearly with $v_R$,
then grows as $v^2_R$ for $v_R \gg 2 \sqrt{\alpha / \beta}\, v^2_u / |z_j|$, corresponding
to $x_j \simeq z_j / \alpha$. This allows one to classify the 8 different solutions
as follows: the 4 solutions with $x_3 = x^-_3$ are characterized by a constant
value of the lightest right-handed neutrino mass, $M_1 \simeq 10^5$ GeV
(the ``type I value'' of $M_1$) over the considered range
of values for $v_R$. Among these 4 solutions, the 2 solutions with $x_2 = x^-_2$
also have a constant value of $M_2$ above $v_R \sim 10^{13}$ GeV.
The 2 solutions with $x_3 = x^+_3$ and $x_2 = x^-_2$ are characterized
by a constant value of $M_1$ above $v_R \sim 10^{13}$ GeV,
$M_1 = 5 \times 10^9$ GeV, and by a crossing of $M_1$ and $M_2$
close to $v_R = 3 \times 10^{12}$ GeV. Finally, the 2 solutions with
$x_3 = x^+_3$ and $x_2 = x^+_2$ are characterized by a rising $M_1$.

As can be seen from Fig.~\ref{fig:Mi_ref}, the perturbativity
constraint forbids large values of $v_R$ in all solutions but $(-,-,-)$.
The associated upper limit on $v_R$ ranges from $3 \times 10^{14}$
to $3 \times 10^{15}$ GeV (except for solution $(-,-,-)$ for which there is
no such constraint on $v_R$), which excludes a breaking of $B-L$ above,
at or just below the GUT scale. However, as discussed below, this conclusion
strongly depends on the value of the ratio $\beta / \alpha$,
assumed here to be $1$. If one requires
in addition the absence of a strong cancellation between the type I and type II
contributions to neutrino masses, the allowed values of $v_R$ are restricted
to a rather small range.

Let us now consider the patterns of right-handed neutrino mixing angles
(Fig.~\ref{fig:Uij_ref}). As in the case of the mass eigenvalues, one can
recognize known limits. The type I and the type II limits are
recovered in the large $v_R$ region of solutions $(-,-,-)$ and $(+,+,+)$,
respectively. The type I limit is characterized by small mixing angles,
very close to the CKM angles, while in the type II limit
where $f \rightarrow M_\nu / v_L$,
the right-handed neutrino mixing angles are given by the PMNS angles
(and since $\theta_{13}$ is close to its present experimental upper bound
in the fit that we used, even $|(U_f)_{13}|$ is relatively large in this limit).
In the small $v_R$ region ($v_R \lesssim 10^{10}$ GeV), the mixing
angles are close to the CKM angles in all solutions.
This can be immediately understood in the $(+,+,+)$ and $(-,-,-)$ cases,
in which, as discussed in Subsection~\ref{subsec:properties}, $f$ tends to
$\pm \sqrt{\beta / \alpha}\, Y_u$
when $4 \alpha \beta \gg |z_3|^2$; in the other cases, $U_f \approx U^T_q$
is a consequence of the hierarchical structure of the Dirac matrix
(see Appendix~\ref{app:analytical}). Some striking features of the mixing
patterns emerge. In solutions $(+,-,-)$ and $(-,-,-)$, the mixing
angles are almost independent of $v_R$.
In the other 6 solutions, the mixing angles evolve from $U_f \approx U^T_q$
at $v_R \lesssim 10^{10}$ GeV to significantly larger values at large $v_R$
(cancellations may occur for some specific values of $v_R$). The $(1,2)$
mixing angle is always of the order of the Cabibbo angle or greater,
except in solutions $(+,+,+)$ and $(-,+,+)$ where a cancellation occurs
close to $v_R = 2 \times 10^{12}$ GeV.

So far we only considered the reference case of a hierarchical
light neutrino mass spectrum with $m_1 = 10^{-3}$ eV, $\beta = \alpha$
and no CP violation beyond the CKM case.
It is interesting to see how different input parameters would
affect the results of Figs.~\ref{fig:Mi_ref} and \ref{fig:Uij_ref}. 
In the following, we briefly discuss the impact on the right-handed neutrino
mass spectrum of the ratio $\beta / \alpha$, of the high energy phases
and of the type of the light neutrino mass hierarchy.

\begin{figure}
\begin{center}
\includegraphics*[height=4.5cm]{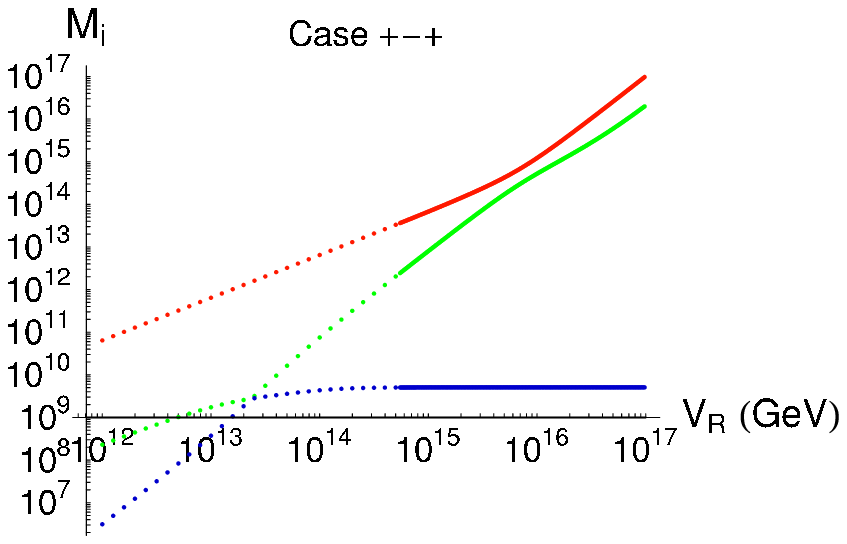}
\hskip 1cm
\includegraphics*[height=4.5cm]{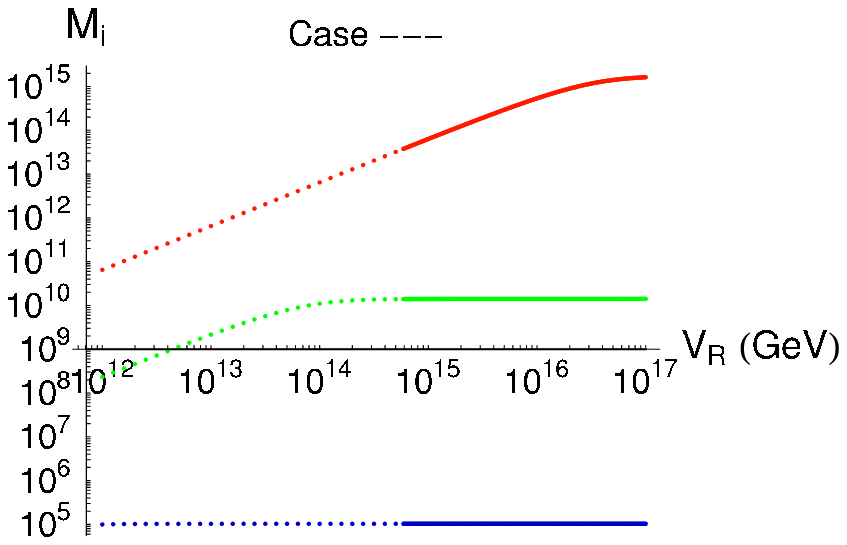}
\vskip .5cm
\caption{Effect of $\beta \neq \alpha$ on the right-handed neutrino masses.
The input parameters are the same as in Fig.~\ref{fig:Mi_ref}, except
$\beta / \alpha = 0.01$.}
\label{fig:Mi_beta_alpha}
\end{center}
\end{figure}

Let us first consider the effect of $\beta \neq \alpha$.
As can be seen by comparing Figs.~\ref{fig:Mi_ref} and~\ref{fig:Mi_beta_alpha},
taking $\beta \neq \alpha$
does not change the general shape of the solutions, but amounts to shift
the curves $M_i = M_i (v_R)$ along the horizontal axis according
to $v_R \rightarrow \sqrt{\alpha / \beta}\, v_R$
(an analogous statement can be made about the curves
$(U_f)_{ij} = (U_f)_{ij} (v_R)$).
For instance, in solution $(-,-,-)$,
the type I limit is reached at larger $v_R$ values for $\beta / \alpha = 0.01$
than for $\beta = \alpha$ (see the right panel of Fig.~\ref{fig:Mi_beta_alpha}).
Nevertheless the values of the $M_i$ corresponding to a plateau do not depend on
$\beta / \alpha$. In particular, in the four solutions characterized by $x_3 = x^-_3$,
one has $M_1 \simeq 10^5$ GeV over the considered range of values
for $v_R$, irrespective of the value of $\beta / \alpha$.
Finally, the value of $\beta / \alpha$ has a strong impact on the allowed
range of values for $v_R$: as can be seen in the left panel of
Fig.~\ref{fig:Mi_beta_alpha}, the perturbativity constraint is more
easily satisfied for large values of $v_R$ when $\beta / \alpha \ll 1$.
This is due to the fact that the asymptotic value of $|f_{33}|$ in the small $v_R$
region, $\sqrt{\beta / \alpha}\, |(Y_\nu)_{33}|$, is proportional to $\sqrt{\beta / \alpha}$.
Conversely, the case $\beta / \alpha \gg 1$ is excluded
because the perturbativity constraint $|f_{33}| < 1$
is never satisfied, except in the type I limit of solution $(-,-,-)$.
Therefore, perturbativity constrains the $SU(2)_L$ triplet mass
to lie below the $B-L$ brealing scale (which might require a fine-tuning
in the $SU(2)_L$ triplet mass matrix for $v_R \ll M_{GUT}$),
except in the type I limit.

\begin{figure}
\begin{center}
\includegraphics*[height=4.5cm]{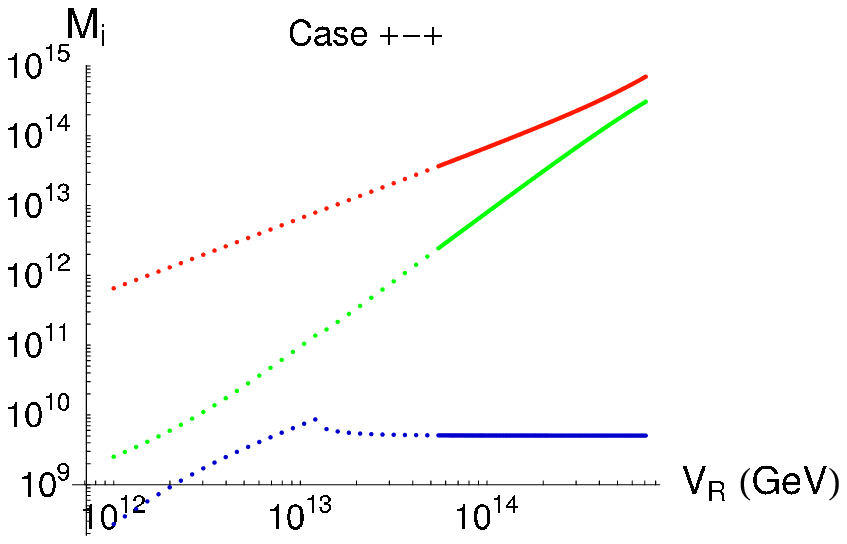}
\hskip 1cm
\includegraphics*[height=4.5cm]{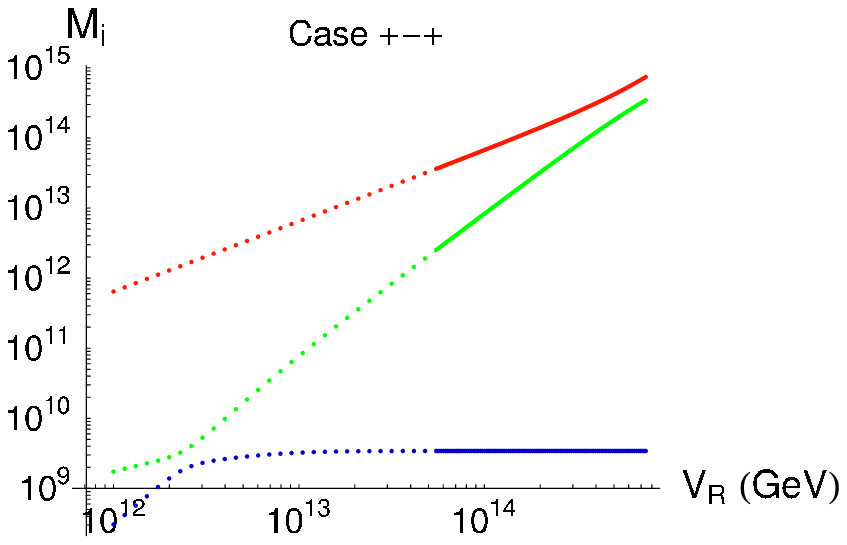}
\vskip .5cm
\caption{Effect of high-energy phases on the right-handed neutrino masses.
The input parameters are the same as in Fig.~\ref{fig:Mi_ref}, except
$\Phi^u_2 = \pi / 4$ (left panel), $\Phi^d_1 = \pi / 4$ (right panel).}
\vskip .5cm
\label{fig:Mi_phases}
\end{center}
\end{figure}

The effect of input CP-violating phases other than the CKM phase
on the right-handed neutrino masses is illustrated in Fig.~\ref{fig:Mi_phases}. 
In general, the presence of these phases only slightly affects the shape
of the solutions, except in regions where
a crossing of two mass eigenvalues occurs. Indeed, phases
can lift isolated degeneracies between two eigenvalues (the curves repel
one another instead of crossing), thus sensibly modifying the shape
of the solution\footnote{The opposite situation can also happen,
i.e. input CP-violating phases can induce a crossing between
two mass eigenvalues in cases where the corresponding curves
do not intersect in the absence of phases.}.
An example of this effect is shown in Fig.~\ref{fig:Mi_phases},
where the solution $(+,-,+)$ is displayed for two different choices
of a non-zero high-energy phase, $\Phi^u_2 = \pi / 4$ (left panel)
and $\Phi^d_1 = \pi / 4$ (right panel). These plots are to be compared
with the corresponding plot in Fig.~\ref{fig:Mi_ref}, where a crossing
between $M_1$ and $M_2$ occurs at $v_R \simeq 3 \times 10^{12}$ GeV.
As for the right-handed neutrino mixing angles $(U_f)_{ij}$,
they are even more sensitive to input CP-violating phases than the $M_i$.

\begin{figure}
\begin{center}
\includegraphics*[height=4.5cm]{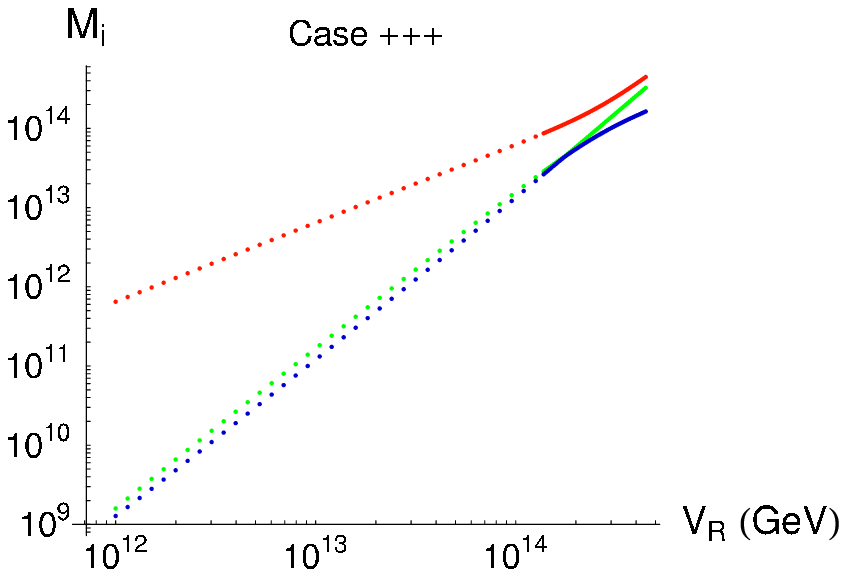}
\hskip 1cm
\includegraphics*[height=4.5cm]{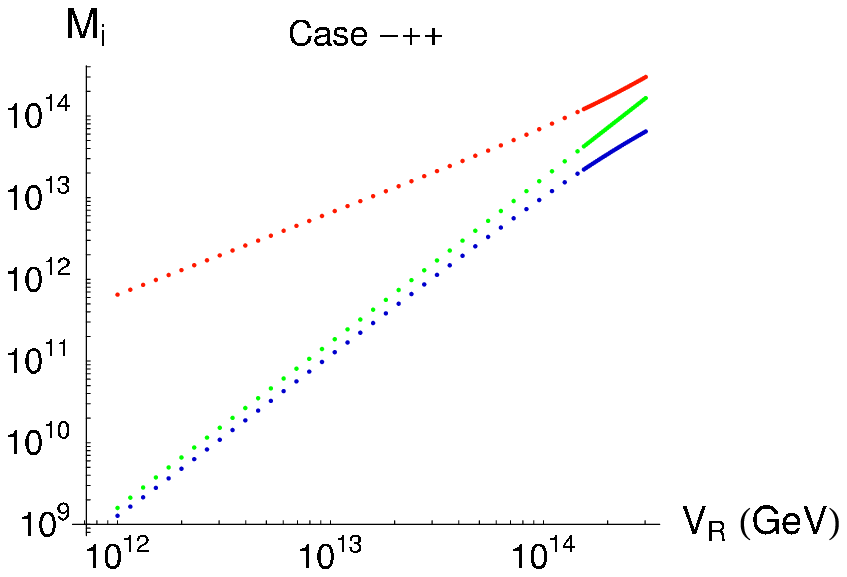}
\vskip .5cm
\caption{Effect of the
light neutrino mass hierarchy on the right-handed neutrino masses.
The input parameters are the same as in Fig.~\ref{fig:Mi_ref}, except
that the light neutrino mass hierarchy is inverted, with $m_3 = 10^{-3}$ eV
and opposite CP parities for $m_1$ and $m_2$.}
\label{fig:Mi_inverted_hierarchy}
\end{center}
\end{figure}

Finally, the right-handed neutrino mass and mixing patterns also depend on
the light neutrino parameters that serve as an input in the reconstruction
procedure, some of which are still unknown ($m_1$, $\mbox{sign}\, (\Delta m^2_{32})$,
$\theta_{13}$, $\delta$ and the two Majorana phases contained in $P_\nu$).
It has already been shown in the type I case that particular values
of these parameters can drastically modify the pattern of right-handed
neutrino masses obtained in the generic case~\cite{AFS03}.
It would be very interesting to investigate such effects in the case
considered here; however, a general study of the dependence
of the 8 right-handed neutrino spectra on the light neutrino mass
parameters is beyond the scope of this paper. We just show in passing
(Fig.~\ref{fig:Mi_inverted_hierarchy}) the impact of the type of the light neutrino
mass hierarchy on right-handed neutrino masses,
for the two solutions where the effect is the most significant.

\section{Leptogenesis}
\label{sec:leptogenesis}

In the previous section, we showed on a particular $SO(10)$ example
that the spectrum of possibilities to account for the experimental neutrino
data in the presence of both type I and type II seesaw mechanisms is very rich.
This has of course important implications for phenomena in which the presence
of right-handed neutrinos and/or of a heavy $SU(2)_L$ triplet plays a role,
such as leptogenesis and, in supersymmetric theories, lepton flavour violation.

In this section, we show that taking into account both seesaw contributions
to neutrino masses opens up new possibilities for successful leptogenesis
in $SO(10)$ GUTs. Since the model we consider is not fully realistic
as it leads to wrong mass relations between charged fermions, we do not
undertake a full study of leptogenesis including washout effects, but consider
solely the value of the CP asymmetry.
We do not try either to maximize the asymmetry by playing with all input
parameters (in particular, we stick to a hierarchical light neutrino spectrum
with $m_1 = 10^{-3}$ eV and the best fit values (\ref{eq:masses_fit})
and (\ref{eq:angles_fit}) for the oscillation parameters), but we restrict
our attention to the impact of the input CP-violating phases.

In the scenario we are considering, it is natural to assume that the lightest
right-handed neutrino is lighter than the $SU(2)_L$ triplet. Indeed, while the
perturbativity constraint discussed in Subsection~\ref{subsec:spectra}
requires $M_{\Delta_L} \lesssim v_R$,
$M_1$ lies several orders of magnitude below $v_R$.
Thus, one can safely assume
that $M_1 \ll M_{\Delta_L}$, in which case the lepton asymmetry is dominantly
generated in out-of-equilibrium decays of the lightest right-handed (s)neutrino.
The CP asymmetry $\epsilon_{N_1} \equiv \left[ \Gamma (N_1 \rightarrow l H)
- \Gamma (N_1 \rightarrow \bar l H^\star) \right]  / \left[ \Gamma (N_1 \rightarrow l H)
+ \Gamma (N_1 \rightarrow \bar l H ^\star) \right]$ receives two contributions:
the standard type I contribution $\epsilon^I_{N_1}$~\cite{FY86,epsilonI},
and a contribution from a vertex diagram containing a virtual triplet,
$\epsilon^{II}_{N_1}$~\cite{epsilonII,HS03}.
In the case $M_1 \ll M_{2,3}$ which is relevant here,
they can be written as~\cite{HS03,AK04}:
\beq
  \epsilon^{I (II)}_{N_1}\ =\ \frac{3}{8 \pi}\
  \frac{\sum_{k, l} \mbox{Im} \left[ Y_{1k} Y_{1l}\,
  (M^{I (II)}_\nu)^\star_{kl} \right]}{(Y Y^\dagger)_{11}}\ \frac{M_1}{v^2}\ ,
\label{eq:epsilon_I_II}
\eeq
where $M^I_\nu \equiv - \beta\, Y f^{-1} Y$ and $M^{II}_\nu \equiv \alpha f$ are the
type I and type II contributions to the neutrino mass matrix, respectively.
The total CP asymmetry in $N_1$ decays then reads:
\beq
  \epsilon_{N_1}\ =\  \epsilon^I_{N_1} +  \epsilon^{II}_{N_1}\ =\ \frac{3}{8 \pi}\
  \frac{\sum_{k, l} \mbox{Im} \left[ Y_{1k} Y_{1l}\,
  (M_\nu)^\star_{kl} \right]}
  {(Y Y^\dagger)_{11}}\ \frac{M_1}{v^2}\ .
\label{eq:epsilon_total}
\eeq
In Eqs.~(\ref{eq:epsilon_I_II}) and (\ref{eq:epsilon_total}), the Dirac couplings
are expressed in the basis of charged lepton and right-handed neutrino
mass eigenstates, i.e. $Y_{1k} \equiv (U^\dagger_f Y_\nu)_{1k}$.
Besides its obvious dependence on the light neutrino mass matrix
and on the phases it contains,
the CP asymmetry depends on the considered solution for the
matrix $f$ and on the input parameters (in particular on the phases)
through their influence on the values of $M_1$ and of the right-handed
neutrino mixing angles $(U_f)_{i1}$ ($i = 1,2,3$). 

The final baryon asymmetry is given by:
\beq
  \frac{n_B}{s}\ =\ - 1.48 \times 10^{-3}\, \eta\, \epsilon_{N_1}\ ,
\label{eq:n_B_s}
\eeq
where $\eta$ is an efficiency factor that takes into account the initial
population of right-handed (s)neutrinos, the out-of-equilibrium condition
for their decays, and the subsequent dilution of the generated lepton asymmetry
by wash-out processes~\cite{GNRRS03}. For leptogenesis to be successful,
Eq.~(\ref{eq:n_B_s}) should reproduce the observed baryon-to-entropy ratio
$n_B / s = (8.7 \pm 0.3) \times 10^{-11}$~\cite{WMAP}.
Detailed studies of thermal leptogenesis in the type I case
(see e.g. Refs.~\cite{BDP02,GNRRS03}) have shown
that $\eta \geq 0.1$ over a significant portion of the parameter space;
therefore thermal leptogenesis can succesfully generate the observed
cosmological baryon asymmetry for $|\epsilon_{N_1}| \sim 10^{-6}$
(or even for $|\epsilon_{N_1}| \sim \mbox{few} \times 10^{-7}$ in the case
of a thermal initial population of $N_1$ / $\tilde N_1$).
Large efficiency factors can also be obtained in the presence of both type I
and type II seesaw mechanisms~\cite{HS03}.

\begin{figure}
\begin{center}
\includegraphics*[height=4.5cm]{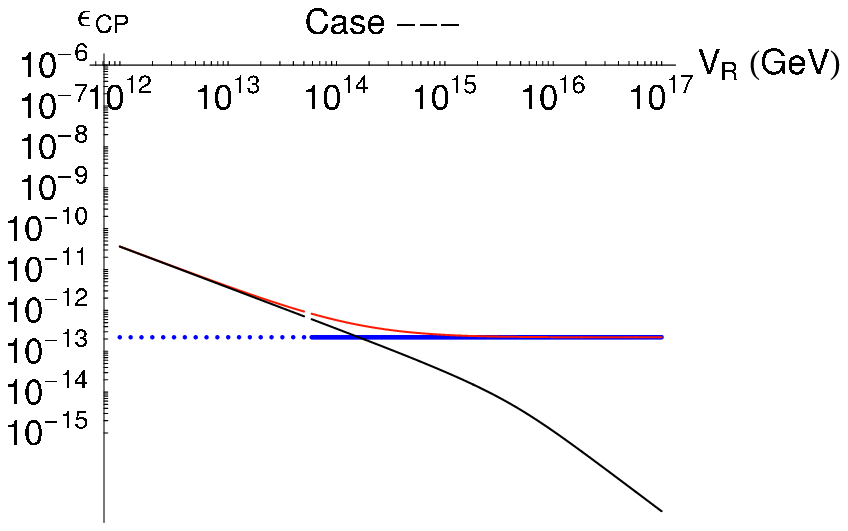}
\hskip 1cm
\includegraphics*[height=4.5cm]{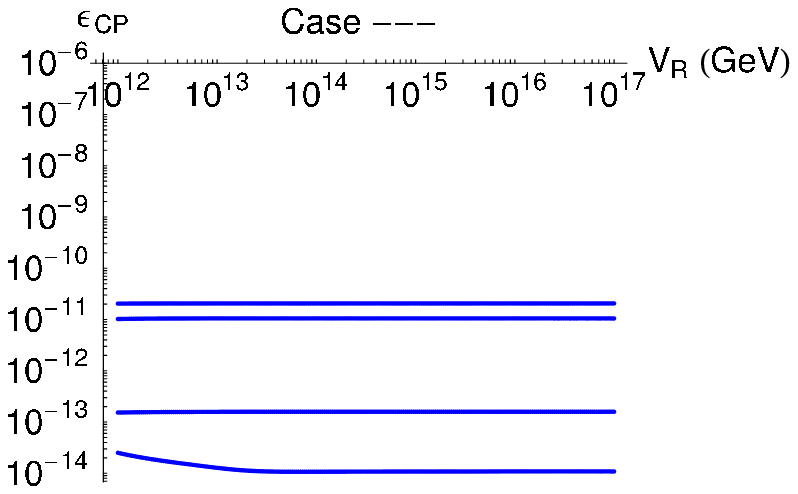}
\vskip .5cm
\caption{CP asymmetry $\epsilon_1$ as a function of $v_R$ for the solution $(-,-,-)$
in the case of a hierarchical light neutrino mass spectrum with $m_1 = 10^{-3}$ eV,
$\beta = \alpha$, and no CP violation beyond the CKM phase (left panel)
or different choices of CP-violating phases (right panel). On the left panel, 
the thin lines correspond to the contribution of right-handed neutrinos
(red [grey] curve) and of the heavy triplet (black curve).}
\label{fig:epsilon_---}
\end{center}
\end{figure}

\begin{figure}
\begin{center}
\includegraphics*[height=4.5cm]{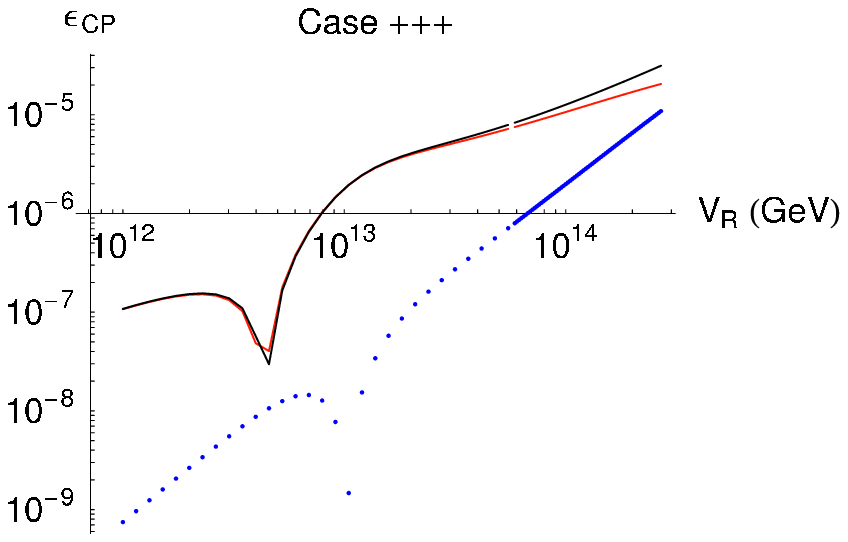}
\hskip 1cm
\includegraphics*[height=4.5cm]{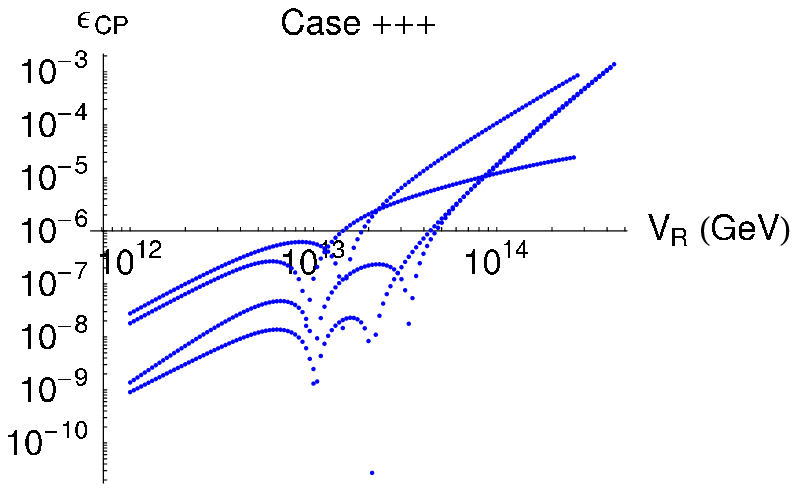}
\vskip .5cm
\caption{Same as Fig.~\ref{fig:epsilon_---} but for the solution $(+,+,+)$.}
\label{fig:epsilon_+++}
\end{center}
\end{figure}

\begin{figure}
\begin{center}
\includegraphics*[height=4cm]{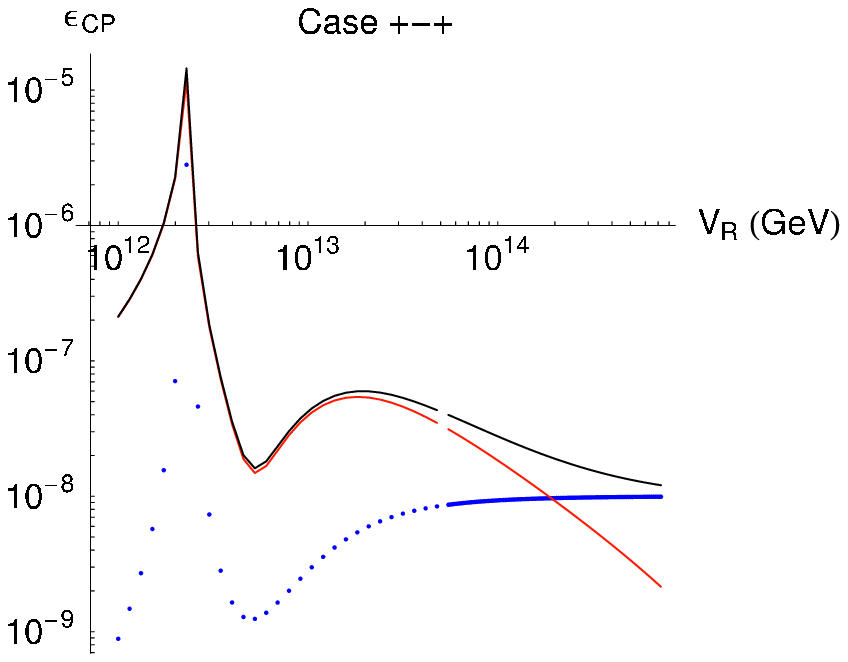}
\hskip .5cm
\includegraphics*[height=4cm]{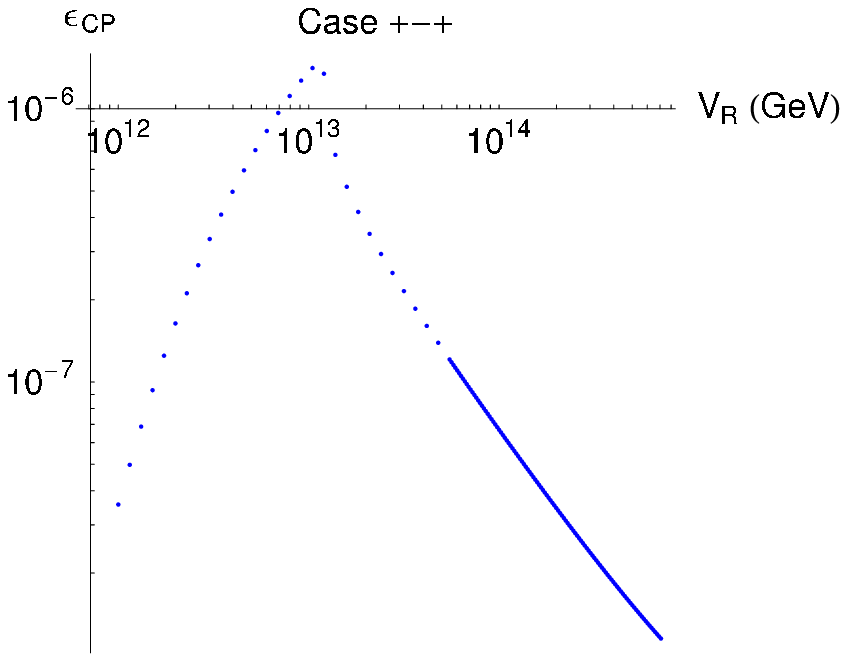}
\hskip .5cm
\includegraphics*[height=4cm]{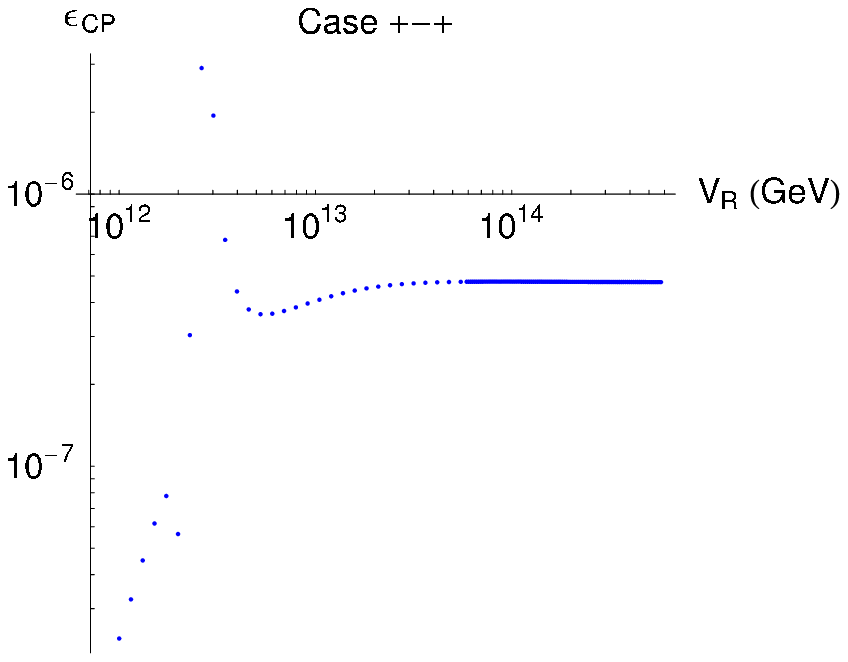}
\vskip .5cm
\caption{Same as Fig.~\ref{fig:epsilon_---} but for the solution $(+,-,+)$,
with no CP violation beyond the CKM phase (left panel),
$\Phi^u_2 = \pi / 4$ (middle panel) and $\Phi^\nu_2 = \pi / 4$ (right panel).}
\label{fig:epsilon_+-+}
\end{center}
\end{figure}

Figs.~\ref{fig:epsilon_---} to~\ref{fig:epsilon_+-+} show the absolute value
of the CP asymmetry in $N_1$ decays as a function of $v_R$, for three
representative solutions $(-,-,-)$, $(+,+,+)$ and $(+,-,+)$.
Before commenting on these results, let us note that an upper bound
on $\epsilon_{N_1}$ can be derived
from Eq.~(\ref{eq:epsilon_total})~\cite{HS03,AK04}:
\beq
  |\epsilon_{N_1}|\ \leq\ \epsilon^{max}_{N_1}
  \equiv \frac{3}{8 \pi}\ \frac{M_1 m_{max}}{v^2}\
  \simeq\ 2 \times 10^{-7} \left( \frac{M_1}{10^9\, \mbox{GeV}} \right)
  \left( \frac{m_{max}}{0.05\, \mbox{eV}} \right) , 
\label{eq:epsilon_max}  
\eeq
where $m_{max} \equiv \max\, (m_1, m_2, m_3)$.
From this one can already conclude that, for a generic\footnote{It has been 
shown in Ref.~\cite{AFS03}, in the context of the type I seesaw mechanism
with a strongly hierarchical Dirac mass matrix, that for some special values
of the light neutrino mass parameters,
the right-handed neutrino mass matrix exhibits a pseudo-Dirac structure,
making it possible to generate the observed baryon asymmetry through
resonant leptogenesis.} hierarchical light neutrino mass spectrum,
the four solutions characterized by $x_3 = x^-_3$,
which give $M_1 \sim 10^5$ GeV for all values of $v_R$,
fail to generate the observed baryon asymmetry from $N_1$ decays
(we comment at the end of this section on possible flavour effects).
This is confirmed by Fig.~\ref{fig:epsilon_---}, which shows that,
depending on the values of the CP-violating phases, $|\epsilon_{N_1}|$
ranges from $10^{-14}$ to $2 \times 10^{-11}$ in the $(-,-,-)$ solution.
As expected,  the CP asymmetry is dominated
by the type I contribution for large values of $v_R$, while the type I
and the type II contributions become comparable and start cancelling
each other below $v_R \sim 10^{14}$ GeV. The most noticeable fact here
is that $\epsilon_{N_1}$ (like $M_1$) stays constant at its type I value even
far away from the type I limit.

The four solutions characterized by $x_3 = x^+_3$,
which for $v_R > 10^{13}$ GeV give either $M_1 \sim 5 \times 10^9$ GeV
(case $x_2 = x^-_2$) or $M_1 > 10^{10}$ GeV (case $x_2 = x^+_2$),
look much more promising.
Indeed, solutions $(+,+,+)$ and $(-,+,+)$ (case $x_2 = x^+_2$) yield
large values of $\epsilon_{N_1}$, even in the absence of other sources
of CP violation than the CKM phase (see Fig.~\ref{fig:epsilon_+++}).
However, the effective mass parameter
$\tilde m_1 \equiv (Y Y^\dagger)_{11} v^2 / M_1$, which controls
the out-of-equilibrium condition and the wash-out due to inverse $N_1$
decays, tends to be rather large (typically $\tilde m_1 \sim 10^{-2}$ eV).
The corresponding suppression of the final baryon asymmetry
can be compensated for by larger values of $\epsilon_{N_1}$,
but at the price of a heavier right-handed neutrino: one typically has
$|\epsilon_{N_1}| > 10^{-5}$ for $M_1 \gtrsim 10^{11}$ GeV.
Such values of $M_1$ are in conflict with the upper limit on the reheating
temperature from gravitino overproduction, which depending on the gravitino
mass and decay modes may lie between $10^6$ and $10^{10}$ GeV~\cite{KKM04}.
One may circumvent this problem by invoking a non-thermal mechanism
for producing right-handed (s)neutrinos after inflation, e. g. decays
of the inflaton field~\cite{inflaton}.

Solutions $(+,-,+)$ and $(-,-,+)$ (case $x_2 = x^-_2$) are in principle
better candidates for a successful thermal leptogenesis since they predict
$M_1 \sim 5 \times 10^9$ GeV, a value
that can lead to a sufficient CP asymmetry while being marginally
compatible with the gravitino constraint.
As shown by Fig.~\ref{fig:epsilon_+-+}, the CP asymmetry
generally reaches a plateau above $v_R \sim 10^{13}$ GeV,
where depending on the phases it can be as large as $5 \times 10^{-7}$
(interestingly enough, this may solely be due to low-energy
CP-violating phases -- see the right panel of Fig.~\ref{fig:epsilon_+-+}).
Such values of $\epsilon_{N_1}$ could be sufficient for
generating the observed baryon asymmetry, provided that the
wash-out processes are slow enough. However, the effective
mass parameter $\tilde m_1$ tends to be too large, typically
$\tilde m_1 > 10^{-2}$ eV. Larger values of $\epsilon_{N_1}$
can be obtained in the region where a strong cancellation
between the type I and type II contributions to neutrino masses 
occur.  In the left and right panels of Fig.~\ref{fig:epsilon_+-+}, the peak
located at $v_R \approx 3 \times 10^{12}$ GeV is due to a near degeneracy
between $M_1$ and $M_2$; there resonant leptogenesis~\cite{resonant}
becomes possible. In the middle panel, the enhancement of $\epsilon_{N_1}$
around $v_R = 10^{13}$ GeV is not related to any mass degeneracy
and is simply an effect of the phase $\Phi^u_2$. In this case too, the wash-out
of the generated lepton asymmetry is strong ($\tilde m_1 \approx 0.03$ eV).

Before closing this section, let us comment on possible flavour
effects~\cite{BCST00,P05,V05,ADJLR06,NNRR06},
specializing for definiteness to the type I limit of solution $(-,-,-)$.
The relevant quantities are the CP asymmetries
in the decays of one right-handed neutrino flavour
$N_i$ into one charged lepton flavour $l_\alpha$, defined as
$\epsilon^\alpha_i \equiv \left[ \Gamma (N_i \rightarrow l_\alpha H)
- \Gamma (N_i \rightarrow \bar l_\alpha H^\star) \right] /$
$\left[ \Gamma (N_i \rightarrow l H) + \Gamma (N_i \rightarrow \bar l H^\star) \right]$,
as well as the parameters $\tilde m^\alpha_i \equiv |Y_{i \alpha}|^2 v^2 / M_i$,
which control the out-of-equilibrium conditions and the main wash-out processes.
Because of the smallness
of its mass, including flavour effects in the decays of the lightest right-handed
neutrino $N_1$~\cite{ADJLR06,NNRR06} does not improve the situation;
but it has been suggested that decays of the next-to-lightest right-handed
neutrino $N_2$ (whose mass is $M_2 \simeq 2 \times 10^{10}$ GeV here)
might lead to successful leptogenesis, without~\cite{DB05} or with~\cite{V05}
flavour effects. One interesting possibility~\cite{V05} is that $N_2$ decays
generate a large asymmetry in a specific lepton flavour that is only mildly erased
by $N_1$ decays and inverse decays.
Whether this can happen or not depends on the values of the parameters
$\epsilon^\alpha_2$, $\tilde m^\alpha_2$ and $\tilde m^\alpha_1$ which
we give in Table~\ref{tab:flavoured_parameters}, together with the other
flavoured parameters for completeness. In the case considered (hierarchical
light neutrino mass spectrum with $m_1 = 10^{-3}$ eV, $\Phi^\nu_2 = \pi / 4$
and all other CP-violating phases but the CKM phase set to zero), we find
that the lepton asymmetry is essentially generated in the tau flavour;
unfortunately it is small ($\epsilon^\tau_2 = 1.4 \times 10^{-7}$)
and the wash-out by $N_1$ decays turns out to be strong
($\tilde m^\tau_1 = 2.2 \times 10^{-2}$). Different choices of the CP-violating
phases might however improve the situation.

\begin{table}[h]
\begin{center}
\begin{tabular}{|c|c|c|c|}   \hline
{parameter / lepton flavour} & \quad {$\alpha = e$} \quad
& \quad {$\alpha = \mu$} \quad & \quad {$\alpha = \tau$} \quad  \\
\hline \hline
$\epsilon^\alpha_1$ \quad  & \quad $2.7 \times 10^{-13}$ \quad
& \quad $- 6.0 \times 10^{-12}$ \quad & \quad $- 1.5 \times 10^{-11}$ \quad \\
\hline
$\epsilon^\alpha_2$ \quad &  \quad $5.6 \times 10^{-11}$ \quad
& \quad $- 1.5 \times 10^{-9}$ \quad & \quad $1.4 \times 10^{-7}$ \quad  \\
\hline
$\epsilon^\alpha_3$ \quad & \quad $- 1.8 \times 10^{-14}$ \quad
& \quad $5.0 \times 10^{-13}$ \quad & \quad $- 4.5 \times 10^{-11}$ \quad \\
\hline
$\tilde m^\alpha_1$ \quad &  \quad $3.3 \times 10^{-3}$ eV \quad
& \quad $1.6 \times 10^{-2}$ eV \quad & \quad $2.2 \times 10^{-2}$ eV \quad  \\
\hline
$\tilde m^\alpha_2$ \quad &  \quad $5.9 \times 10^{-4}$ eV \quad
& \quad $1.1 \times 10^{-2}$ eV \quad & \quad $3.5 \times 10^{-2}$ eV \quad  \\
\hline
$\tilde m^\alpha_3$ \quad &  \quad $4.0 \times 10^{-7}$ eV \quad
& \quad $1.1 \times 10^{-5}$ eV \quad & \quad $9.4 \times 10^{-3}$ eV \quad  \\
\hline
\end{tabular}
\caption{Parameters that control flavour effects in leptogenesis in the type I
case (large $v_R$ limit of solution $(-,-,-)$), in the case of a hierarchical
light neutrino mass spectrum with $m_1 = 10^{-3}$ eV,
$\Phi^\nu_2 = \pi / 4$ and all other CP-violating phases but the CKM phase
set to zero.}
\label{tab:flavoured_parameters}
\end{center}
\end{table}

The above discussion shows that taking into account both the type I
and the type II seesaw contributions to neutrino masses opens up
new possibilities for successful leptogenesis in $SO(10)$ GUTs,
even though, for the specific choice of input parameters made
in this paper, the wash-out processes tend to be too strong.
Different choices for the light neutrino mass parameters,
or different combinations of the high-energy phases,
could resolve this problem.
Let us also recall that the results presented in this section
were obtained using the mass relations (\ref{eq:M_10}),
which need to be corrected. The inclusion of corrections leading to
realistic charged fermion mass matrices, e.g. from the $\bf \overline{126}$
Higgs representation, is not expected to alter the gross qualitative features
of the right-handed neutrino mass spectrum, but might modify the numerical
values of $M_1$ and of the right-handed neutrino mixing angles, hence
the predictions for leptogenesis.


\section{Lepton flavour violation}
\label{sec:LFV}

In supersymmetric extensions of the Standard Model,
lepton flavour violating (LFV) processes such as
the charged lepton radiative decays $l_j \rightarrow l_i \gamma$
arise from loop diagrams involving sleptons and charginos/neutralinos.
The relevant flavour-violating parameters are the off-diagonal entries
of the slepton soft supersymmetry breaking mass matrices
$(m^2_{\tilde L})_{ij}$, $(m^2_{\tilde e_R})_{ij}$
and $(m^{e\, 2}_{RL})_{ij} \equiv A^e_{ij} v_d$,
expressed in the flavour basis defined by the charged lepton
mass eigenstates.

If the supersymmetry breaking mechanism is flavour blind,
flavour violation in the slepton sector arises from radiative corrections
induced by the flavour-violating couplings of heavy states populating
the theory between the Planck scale and the electroweak scale.
Here we must deal with two kinds of such couplings\footnote{We do not
consider the other sources of lepton flavour violation that can be present
in supersymmetric GUTs, such as the contribution of colour triplets~\cite{BHS95}
or the contribution of the $SU(2)_R$ triplet whose vev is responsible
for right-handed neutrino masses, since they are model dependent.
By contrast the right-handed neutrino and (assuming that $M_{\Delta_L}$
is known) the $SU(2)_L$ triplet contributions can be computed
once the couplings $f_{ij}$ have been reconstructed.}:
the couplings of the right-handed neutrinos~\cite{BM86,HMTYY95,LFVseesaw},
$Y_{ki}$ (where $Y \equiv U^\dagger_f Y_\nu$), and the couplings
of the heavy $SU(2)_L$ triplet~\cite{R02}, $f_{ij}$. Integrating the
one-loop renormalization group equations in the lowest approximation,
one obtains the following expressions for the flavour-violating (off-diagonal)
entries of the soft supersymmetry breaking slepton mass matrices:
\beq
  (m^2_{\tilde L})_{ij}\ \simeq\ -\, \frac{3 m^2_0 + A^2_0}{8 \pi^2}\
  C_{ij}\ , \qquad (m^2_{\tilde e_R})_{ij}\ \simeq\ 0\ ,
  \qquad A^e_{ij}\ \simeq\ -\, \frac{3}{8 \pi^2}\,
  A_0 y_{e_i} C_{ij}\ ,
\eeq
where the coefficients $C_{ij}$ encapsulate the dependence
on the seesaw parameters:
\beq
  C_{ij}\ \equiv\
  \sum_k Y^\star_{ki} Y_{kj}\, \ln \! \left( \frac{M_U}{M_k} \right)
  + 3\, (f f^\dagger)_{ij}\, \ln \! \left( \frac{M_U}{M_{\Delta_L}} \right)\ .
\eeq
Here $M_U$ is the scale at which universality among soft supersymmetry
breaking parameters (at least in the slepton and Higgs sector) is assumed.
In the following, we take $M_U = 10^{17}$ GeV, close to the Landau pole
$\Lambda_{10}$ where the theory becomes non perturbative.
Neglecting the smaller contribution of the flavour-violating $A$-term
and working in the mass insertion approximation,
one can schematically write the branching ratio for
$l_j \rightarrow l_i \gamma$ as:
\beq
  \frac{\mbox{BR}\, (l_j \rightarrow l_i \gamma)}
  {\mbox{BR}\, (l_j \rightarrow l_i \bar \nu_i \nu_j)}\
  \propto\ \frac{|(m^2_{\tilde L})_{ij}|^2}{\bar m^8_{\tilde L}}\
  \tan^2 \beta\ F_{Susy}\ ,
\eeq
where $\bar m^2_{\tilde L}$ is the average slepton doublet mass,
and $F_{Susy}$ is a function of the supersymmetric mass parameters
and of $\tan \beta$. The experimental upper limits
$\mbox{BR}\, (\mu \rightarrow e \gamma) < 1.2 \times 10^{-11}$~\cite{mueg} and
$\mbox{BR}\, (\tau \rightarrow \mu \gamma) < 6.8 \times 10^{-8}$~\cite{taumug}
can then be translated into upper bounds on the $C_{12}$
and $C_{23}$ coefficients as a function of the superpartner masses
and of $\tan \beta$~\cite{LMS01}. If we require that the mSUGRA
parameters $m_0$ and $M_{1/2}$ do not exceed $\sim 1$ TeV,
then from Fig.~3 of Ref.~\cite{MS05} we can read the approximate
upper bounds\footnote{More precisely, for $\tan \beta = 10$, one has
$|C_{12}| < 0.1$ (resp. $|C_{23}| < 20$) for $M_1 < 300$ GeV and
$400\, \mbox{GeV} \lesssim \bar m_{\tilde e_R} \lesssim 1$ TeV if $A_0 = 0$,
and for $M_1 \lesssim 500$ GeV and $\bar m_{\tilde e_R} \lesssim 1$ TeV
if $A_0 = m_0 + M_{1/2}$, where $M_1$ is the bino mass and
$\bar m_{\tilde e_R}$ is the average slepton singlet mass.}
$|C_{12}| \lesssim 0.1$ and $|C_{23}| \lesssim 10$
for a benchmark value of $\tan \beta = 10$. For different values of $\tan \beta$,
the upper bounds approximately scale as $10 / \tan \beta$.

\begin{figure}
\begin{center}
\includegraphics*[height=4.5cm]{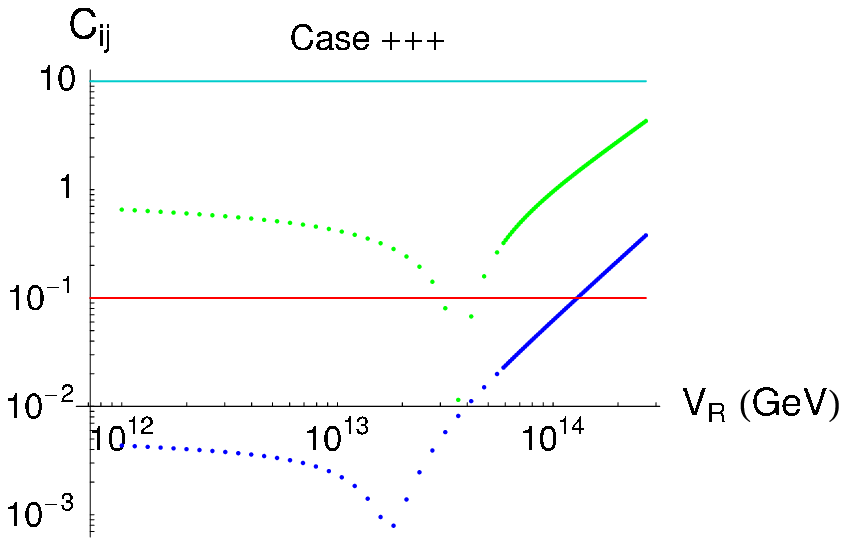}
\hskip 1cm
\includegraphics*[height=4.5cm]{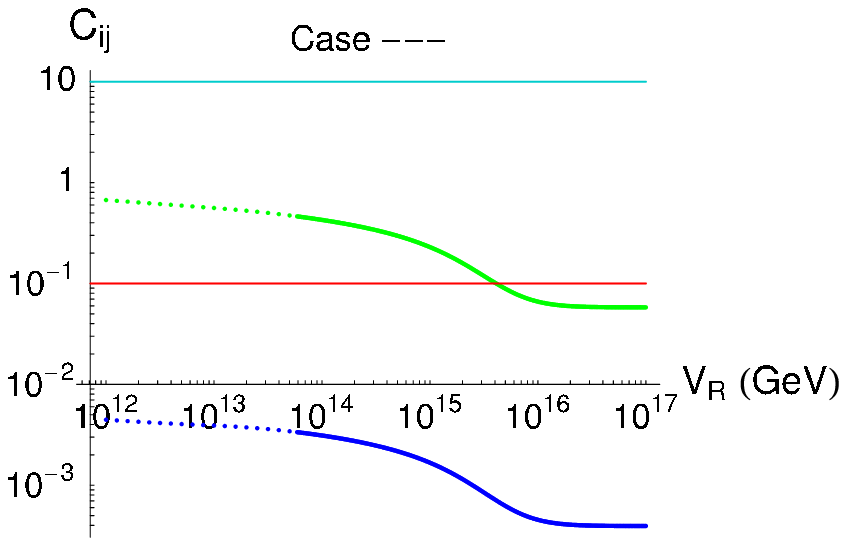}
\vskip .5cm
\caption{Coefficients $C_{12}$ and $C_{23}$ as a function of $v_R$
for the solutions $(+,+,+)$ and $(-,-,-)$ in the case of a hierarchical
light neutrino mass spectrum with $m_1 = 10^{-3}$ eV, $\beta = \alpha$,
and no CP violation beyond the CKM phase.
The green [light grey] curve corresponds to $|C_{23}|$, and the blue [black]
curve to $|C_{12}|$. The horizontal lines indicate the ``experimental'' constraints
$|C_{23}| < 10$ and $|C_{12}| < 0.1$ (see text).}
\label{fig:Cij_ref}
\end{center}
\end{figure}

In Fig.~\ref{fig:Cij_ref}, we compare the values of the coefficients
$|C_{12}|$ and $|C_{23}|$ in solutions $(+,+,+)$ and $(-,-,-)$ with
the ``experimental constraints'' $|C_{12}| < 0.1$ and $|C_{23}| < 10$,
assuming $\beta = \alpha$ and $M_{\Delta_L} = v_R$.
The plot in the left panel of Fig.~\ref{fig:Cij_ref} is representative of all
solutions but $(+,-,-)$ and $(-,-,-)$. One finds that
$\mbox{BR}\, (\tau \rightarrow \mu \gamma)$
lies below the experimental
constraint for all allowed values of $v_R$ (unless $\tan \beta$ is large
and/or some superpartners are light), but it could be accessible
to future experiments for $v_R \gtrsim 10^{14}$ GeV.
The decay $\mu \rightarrow e \gamma$ is much closer to
its present experimental upper limit for larger values of $v_R$,
and even exceeds it for $v_R > 10^{14}$ GeV
(with the assumed values of the supersymmetric parameters).
Solutions $(-,-,-)$ and $(+,-,-)$ have a completely different behaviour;
for these solutions both $\mbox{BR}\, (\tau \rightarrow \mu \gamma)$ and
$\mbox{BR}\, (\mu \rightarrow e \gamma)$ are well below their
experimental uper limits for all values of $v_R$.

\begin{figure}
\begin{center}
\includegraphics*[height=4.5cm]{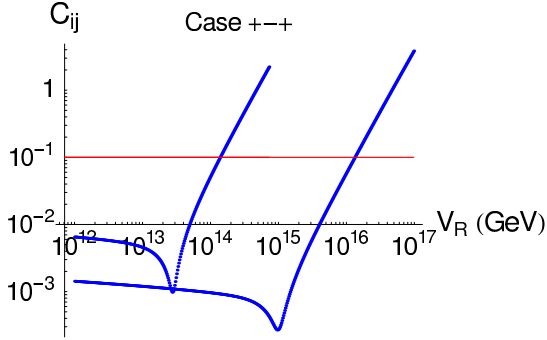}
\vskip .5cm
\caption{Effect of $\beta \neq \alpha$ on $C_{12}$ for the solution $(+,-,+)$.
The input parameters are the same as in Fig.~\ref{fig:Cij_ref}, except that
$\beta / \alpha = 1$ and $M_{\Delta_L} = v_R$ for the left curve, while
$\beta / \alpha = 0.01$ and $M_{\Delta_L} = 0.1 v_R$ for the right curve.}
\label{fig:Cij_beta_alpha}
\end{center}
\end{figure}

We have checked numerically that, except for the large $v_R$ regions
of solutions $(-,-,-)$ and $(+,-,-)$, the type II contribution always dominates
in $C_{12}$ and $C_{23}$ (at least for $\beta \sim \alpha$).
This can easily be understood by noting that, due to the relation
$Y_\nu = Y_u$, the type I contribution is suppressed by the small CKM angles.
The coefficients $C_{ij}$ thus essentially reflect the structure of the matrix $f$,
and like the matrix $f$, they have a strong sensitivity to the CP-violating phases
and to the ratio $\beta / \alpha$.
The impact of $\beta / \alpha$ is shown in Fig.~\ref{fig:Cij_beta_alpha}.
One can see that, for a fixed value of $v_R$, a lower ratio $\beta / \alpha$
results in reduced LFV rates.

To summarize, in the considered class of $SO(10)$ models, LFV processes
are dominated by the type II contribution in most of the parameter space
of the 8 solutions. The predictions lie significantly below the experimental
upper limits, except for the large $v_R$
region of most solutions, where depending on  the values
of the supersymmetric parameters $\mu \rightarrow e \gamma$
can even exceed its present upper limit. The expected improvement
of the experimental sensitivity to LFV processes will strongly constrain
this region.


\section{Conclusions}
\label{sec:conclusions}

The procedure presented in this paper for reconstructing the matrix $f$
of the left-right symmetric seesaw mechanism from the light neutrino
mass parameters can be applied in any theory with an underlying
left-right symmetry which predicts (or at least constrains) the Dirac
mass matrix.
The 8 solutions for the $f_{ij}$ couplings can then be used to study
a number of issues in which the presence of right-handed neutrinos
or heavy $SU(2)_{L (R)}$ triplets (or other heavy states embedded
in the same GUT representation) plays a role, such as leptogenesis,
lepton flavour violation, electric dipole moments of  charged leptons,
or proton decay~\cite{BPW98}. Some of these processes
(e.g. $\mu \rightarrow e \gamma$) put strong constraints
on the 8 solutions, and might even exclude some of them.
The reconstruction procedure can also be used as a tool
to investigate the flavour structure of the right-handed neutrino
mass matrix and to make progress in the quest for a flavour theory.

In this paper, we applied the procedure to a particular class
of supersymmetric $SO(10)$ models with two $\bf 10$-dimensional
and a pair of $\bf 126 \oplus \overline{126}$
representations in the Higgs sector. We found a large variety of
right-handed neutrino spectra compatible with the observed neutrino
data, opening new possibilities for successful leptogenesis
in $SO(10)$ GUTs.
We also studied LFV processes in these models and found large
triplet contributions in most solutions,
especially in the region of large $v_R$ values.
As a byproduct of our study, we found interesting constraints
on the breaking scale of the $B-L$ symmetry,
hence on the masses of the heavy states which play a role
in issues such as gauge coupling unification. 
In particular, for $\beta \sim \alpha$,
the perturbativity constraint excludes values of $v_R$ above a few
$(10^{14} - 10^{15})$ GeV, depending on the solution
(except for the $(-,-,-)$ solution). LFV processes further restrict
the range of allowed values for $v_R$ as a function of
the supersymmetric parameters. Also, it is interesting to note
that the region of values for $v_R$ that is disfavoured by fine-tuning
arguments is the one in which gauge coupling unification
is problematic.

As mentioned earlier, cancellations between the type I and type II
contributions to neutrino masses, rather than being accidental,
could be due to some (broken) symmetry ensuring a proportionality
relation between the right-handed neutrino
and triplet couplings, $f \propto Y_\nu$. This would be particularly
relevant for some interesting possibilities for leptogenesis that occur
in the fine-tuned region, or close to it. Interestingly enough, such a
proportionality relation automatically follows from the embedding
of the $SO(10)$ model
in an $E_6$ GUT with a $\bf 27_H \oplus \overline{27}_H$ and a
$\bf 351' \oplus \overline{351}'$ pairs
in the Higgs sector, in such a way that the $\bf 351'$ representation
contains the $\bf \overline{126}$, $\bf 54$ and $\bf 10_2$ representations
of $SO(10)$, while $\bf 27_H$ contains $\bf 10_1$. With this embedding,
both couplings $\bf 16_i 16_j 10_2$ and $\bf 16_i 16_j \overline{126}$
come from the $\bf 27_i 27_j 351'$ superpotential term. Assuming
that the doublet that couples to up quarks in $\bf 10_1$ does not acquire
a vev, one obtains $f \propto Y_\nu$.
Therefore the presence of a fine-tuning in the seesaw mass formula,
rather than being unnatural, could point to an extended unification in $E_6$.

A few simplifying assumptions were made in this paper. First,
the gauge symmetry breaking aspects of the models (including
the issues of doublet-triplet splitting and gauge coupling unification)
were not taken into account; second, corrections to the ``wrong''
mass relation $M_d = M_e$ were neglected. There are good
reasons to believe that these approximations are justified
at the qualitative level. Indeed, the main features of the right-handed
neutrino spectra are dictated by the strong hierarchical structure
of the Dirac mass matrix, and would not be spoiled by small corrections
to the basic $SO(10)$ mass relations. Nevertheless a more detailed
analysis using realistic mass relations is needed in order to obtain
quantitative predictions, in particular for leptogenesis. Also, a more systematic
scan over the input parameters (most notably the high-energy phases
and the light neutrino masses and mixings) would probably provide
useful information. Work along these lines is in progress.

\vskip0.3cm

\paragraph{Acknowledgments}
We aknowledge useful discussions with M. Frigerio and Th. Hambye.
This work has been supported in part by the RTN European Program 
MRTN-CT-2004-503369 and by the French Program ``Jeunes Chercheurs''
of the Agence Nationale de la Recherche (ANR-05-JCJC-0023).


\begin{appendix}

\section{Perturbativity constraint on the $f_{ij}$ couplings}
\label{app:perturbativity}

In this appendix, we discuss the constraint coming from the requirement
that the couplings $f_{ij}$ remain in the perturbative regime
up to the scale at which the unified gauge coupling $g_{10}$ blows up,
$\Lambda_{10} = M_{GUT} \exp [- \frac{8 \pi^2}{b_{10}\, g^2_{10} (M_{GUT})}]$,
where $b_{10}$ is the $SO(10)$ beta function coefficient
(this Landau pole is due to the presence of a $\bf  126 \oplus \overline{126}$
pair in the Higgs sector, which gives a large contribution to $b_{10}$).

Above $M_{GUT}$, the running of the $f_{ij}$ couplings is governed by
the renormalization group equation:
\beq
  \frac{d f_{ij}}{dt}\ =\
  \frac{1}{2}\, \mbox{Tr} (f^\dagger f) f_{ij} + 15\, (f f^\dagger f)_{ij}
  - \frac{95}{2}\, g^2_{10} f_{ij} + \cdots\ ,
\label{eq:RGE_f_ij}
\eeq
where the dots stand for the contribution of the other superpotential couplings,
which we assume to play a subdominant role in the regime where
the $f_{ij}$'s are large.
Assuming a hierarchy between the eigenvalues, $f_{1, 2} \ll f_3$,
one finds a critical value
$f^{\rm crit.}_3 (M_{GUT}) = \sqrt{(95 - 2 b) / 31}\, g_{10} (M_{GUT})$ above which
$f_3$ diverges before the Landau pole $\Lambda_{10}$ is reached. As an example,
if the Higgs sector contains a $\bf 45$ and a $\bf 54$ representations
in addition to the two $\bf 10$'s and to the $\bf  126 \oplus \overline{126}$ pair,
one obtains $\Lambda_{10} = 1.5 \times 10^{17}$ GeV and
$f^{\rm crit.}_3 (M_{GUT}) \simeq 2$, where we have used
$\alpha_{10} (M_{GUT}) = 1 / 24$ and $M_{GUT} = 2 \times 10^{16}$ GeV.
Since the running of the $f_{ij}$'s below $M_{GUT}$ is much milder
due to the decoupling of the heavy states, we can safely take $f_3 < 1$
as a perturbativity constraint at the scale where the $f_{ij}$'s are determined.

In the above, we have implicitly assumed that $v_R \leq M_{GUT}$.
If this is not the case, $SO(10)$ is broken into $SU(5)$ at the scale $v_R$,
above $M_{GUT}$; as a result the running of the $f_{ij}$'s above $M_{GUT}$
is slower than in the case $v_R \leq M_{GUT}$,
and the Landau pole $\Lambda_{10}$ is shifted towards a larger scale. 
In spite of these differences, $f_3 < 1$ remains a relevant perturbativity
constraint.


\section{Some useful analytical formulae}
\label{app:analytical}

In this appendix, we provide analytical approximations that can
be useful to understand the results of the reconstruction procedure
described in Subsection~\ref{subsec:procedure}.
Although we follow the assumptions of Section~\ref{sec:SO_10},
with $Y_\nu = Y_u$ in the basis of charged lepton mass eigenstates,
the formulae presented below are more generally valid in the case
of a hierarchical Dirac matrix $Y_\nu$, with eigenvalues $y_1 \ll y_2 \ll y_3$
and small mixing angles.

Let us first perform the diagonalization of the matrix $Z$, assuming
for definiteness that the light neutrino mass spectrum is hierarchical
($m_1 < m_2 \ll m_3$).
Using the notations of Subsection~\ref{subsec:models} and
choosing $Y^{1/2} = U^T_q \hat Y^{1/2}_u$, we have:
\beq
  Z\ =\ \hat Y^{-1/2}_u N \hat Y^{-1/2}_u\ ,  \qquad
  N\ \equiv\ (U^\star_q U^\star_l \hat M_\nu U^\dagger_l U^\dagger_q) / m_3\ .
\eeq
Since $m_2 / m_3 \approx 0.2$ and two of the three lepton mixing angles are
large (with $ s_{23} \approx 1 / \sqrt{2}$, $s_{12} \approx 1 / \sqrt{3}$ and
$s_{13} \lesssim 0.2$), the matrix $N$ has a very moderate hierarchy.
We can parametrize it as:
\beq
  N\ =\ \left( \begin{array}{ccc}
  a & b & c  \\
  b & d & e  \\
  c & e & f
  \end{array}  \right)\ ,
\eeq
with $|a|, |b|, |c| \sim m_2 / m_3$, $|d|, |e|, |f| \sim 1$ and $|df - e^2| \sim m_2 / m_3$.
It is convenient to define the quantities:
\begin{eqnarray}
  \Delta_{11}\ \equiv\ df - e^2\ ,
  & \Delta_{12}\ \equiv\ bf - ce\ ,
  & \Delta_{13}\ \equiv\ be - cd\ ,  \\
  \Delta_{22}\ \equiv\ af - c^2\ ,
  & \Delta_{23}\ \equiv\ ae - bc\ ,
  & \Delta_{33}\ \equiv\ ad - b^2\ .
\end{eqnarray}
All $\Delta_{ij}$ are of order $m_2 / m_3$, and
$|\det N| = |a \Delta_{11} - b \Delta_{12} + c \Delta_{13}|
= m_1 m_2 / m^2_3 < (m_2 / m_3)^2$.
The hierarchical structure of the matrix $Z$ is essentially determined
by the up quark Yukawa couplings:
\beq
  Z\ =\ \left( \begin{array}{ccc}
  a / y_u & b / \sqrt{y_u y_c} & c / \sqrt{y_u y_t}  \\
  b / \sqrt{y_u y_c} & d / y_c & e / \sqrt{y_c y_t}  \\
  c / \sqrt{y_u y_t} & e / \sqrt{y_c y_t} & f / y_t
  \end{array}  \right) .
\eeq
The roots of the polynomial equation $\det (Z - z \mathbf{1}) = 0$ are
obviously all distinct, hence $Z$ can be diagonalized by a complex
orthogonal matrix $O_Z$:
\beq
  Z\ =\ O_Z \left( \begin{array}{ccc}
    z_1 & 0 & 0 \\
    0 & z_2 & 0 \\
    0 & 0 & z_3
    \end{array}  \right) O^T_Z\ ,
\eeq
\beq
  z_1\ =\ \frac{\det N}{\Delta_{33}}\, y^{-1}_t\ ,  \qquad
  z_2\ =\ \frac{\Delta_{33}}{a}\, y^{-1}_c\ ,  \qquad
  z_3\ =\ a\, y^{-1}_u\ ,
\label{eq:z_i}
\eeq
\beq
  O_Z\ =\ \left( \begin{array}{ccc}
    \frac{\Delta_{13}}{\Delta_{33}}\, \sqrt{y_u / y_t}
      & - \frac{b}{a}\, \sqrt{y_u / y_c} & 1 \\
    - \frac{\Delta_{23}}{\Delta_{33}}\, \sqrt{y_c / y_t}
      & 1 & \frac{b}{a}\, \sqrt{y_u / y_c} \\
    1 & \frac{\Delta_{23}}{\Delta_{33}}\, \sqrt{y_c / y_t}
      & \frac{c}{a}\, \sqrt{y_u / y_t}
    \end{array}  \right)\ ,
\label{eq:O_Z}
\eeq
where we have ordered the $z_i$ in such a way that $|z_1| < |z_2| < |z_3|$.
In Eqs.~(\ref{eq:z_i}) and~(\ref{eq:O_Z}), the neglected terms are of relative
order $y_u / y_c$, $y_c / y_t$ with respect to the dominant terms.

We can now reconstruct the 8 solutions for the matrix $f$.
For a given choice of ($x_1$, $x_2$, $x_3$),
the matrix $f$ is given by $f = U^T_q \hat Y^{1/2}_u X \hat Y^{1/2}_u U_q$,
with $X = O_Z \mbox{Diag}\, (x_1, x_2, x_3) O^T_Z$. The eigenvalues $f_i$
are obtained by diagonalizing $f$ with a unitary matrix $U_f$,
Eq.~(\ref{eq:diag_f}). Alternatively, one can diagonalize the matrix
$\tilde X \equiv \hat Y^{1/2}_u X \hat Y^{1/2}_u$, which is related to $f$
by $f = U^T_q \tilde X U_q$, and has therefore the same eigenvalues:
\beq
  \tilde X\ =\ U_{\tilde X} \hat f U^T_{\tilde X}\ ,  \qquad
  U_{\tilde X} U^\dagger_{\tilde X} = \mathbf{1}\ ,
\eeq
the unitary matrix that brings $f$ to its diagonal form being given
by $U_f = U^T_q U_{\tilde X}$. It does not seem to be possible
to derive simple analytical formulae for the $f_i$ that would hold for
any value of $\alpha$ and $\beta$. However, one can easily obtain
approximate formulae in the regions of ($\alpha$, $\beta$) values
where the $x_i$ satisfy some hierarchy requirements, as we show below.
Let us first define the following quantities:
\beq
  \bar x_1\ \equiv\ y_t\, x_1\ ,  \qquad  \bar x_2\ \equiv\ y_c\, x_2\ ,
  \qquad  \bar x_3\ \equiv\ y_u\, x_3\ .
\eeq
The $f^2_i$ are given by the roots of the characteristic equation:
\beq
  \det (\tilde X \tilde X^\star - f^2_i \mathbf{1})\
  =\ - f^6_i + C_1 f^4_i - C_2 f^2_i + C_3\ =\ 0, 
\eeq
with (up to subdominant terms of order $y_u / y_c$, $y_c / y_t$ in the
coefficients of the $\bar x_i$ monomials):
\bea
  C_1 & = & \left( \frac{|a|^2 + |b|^2 + |c|^2}{|a|^2} \right)^{\! 2}\, |\bar x_3|^2 
    + \left( \frac{|\Delta_{23}|^2 + |\Delta_{33}|^2}{|\Delta_{33}|^2} \right)^{\! 2}\,
    |\bar x_2|^2 + |\bar x_1|^2  \nn  \\
  & & +\ \left\{ \left( \frac{b^\star}{a^\star} + \frac{c^\star \Delta_{23}}
    {a^\star \Delta_{33}} \right)^{\! 2} \bar x_2 \bar x^\star_3
    + \left( \frac{\Delta^\star_{23}}{\Delta^\star_{33}} \right)^{\! 2} \bar x_1\bar x^\star_2
    + \left( \frac{c^\star}{a^\star} \right)^{\! 2} \bar x_1 \bar x^\star_3
    + \mbox{c.c.} \right\}\, ,  \nn  \\
  C_2 & = & \left( \frac{|\Delta_{13}|^2 + |\Delta_{23}|^2 + |\Delta_{33}|^2}
    {|\Delta_{33}|^2} \right)^{\! 2}\, |\bar x_2 \bar x_3|^2
    + \left( \frac{|a|^2 + |b|^2}{|a|^2} \right)^{\! 2}\, |\bar x_1 \bar x_3|^2
    + |\bar x_1 \bar x_2|^2  \\
  & & +\ \left\{ \left( \frac{\Delta^\star_{23}}{\Delta^\star_{33}}
    + \frac{b \Delta^\star_{13}}{a \Delta^\star_{33}} \right)^{\! 2}
    \bar x_1 \bar x^\star_2 |\bar x_3|^2
    + \left( \frac{\Delta^\star_{13}}{\Delta^\star_{33}} \right)^{\! 2}
    \bar x_1\bar x^\star_3 |\bar x_2|^2
    + \left( \frac{b^\star}{a^\star} \right)^{\! 2} \bar x_2 \bar x^\star_3 |\bar x_1|^2
    + \mbox{c.c.} \right\}\, ,  \nn  \\
  C_3 & = & |\bar x_1 \bar x_2 \bar x_3|^2\ .  \nn
\label{eq:C_i}
\eea
Using the fact that $C_1 = f^2_1 + f^2_2 + f^2_3$,
$C_2 = f^2_1 f^2_2 + f^2_1 f^2_3 + f^2_2 f^2_3$
and $C_3 = f^2_1 f^2_2 f^2_3$,
one immediately sees that, when there is a significant
hierarchy between the $\bar x_i$, the $f_i$ are  given by the
$\bar x_i$ times an order one coefficient. More precisely, one has:
\beq
  f_1 \simeq \sqrt{C_3 / C_2}\ \sim\ \bar x_{min}\ ,  \qquad
  f_2 \simeq \sqrt{C_2 / C_1}\ \sim\ \bar x_{middle}\ ,  \qquad
  f_3 \simeq \sqrt{C_1}\ \sim\ \bar x_{max}\ ,
\label{eq:approximate_f_i}
\eeq
where $\bar x_{max} \equiv \max (|\bar x_1|, |\bar x_2|, |\bar x_3|)$,
$\bar x_{min} \equiv \min (|\bar x_1|, |\bar x_2|, |\bar x_3|)$ and
$\bar x_{middle}$ is the remaining $\bar x_i$.
In the case $|\bar x_1| \ll |\bar x_2| \ll |\bar x_3|$, this reads:
\bea
  & f_1\ \simeq\ \frac{|\Delta_{33}|^2}{|\Delta_{13}|^2
    + |\Delta_{23}|^2 + |\Delta_{33}|^2}\ |\bar x_1|\ ,  \qquad
  f_2\ \simeq\ \frac{|a|^2 (|\Delta_{13}|^2 + |\Delta_{23}|^2 + |\Delta_{33}|^2)}
    {(|a|^2 + |b|^2 + |c|^2) |\Delta_{33}|^2}\ |\bar x_2|\ ,  \nn  \\
  & f_3\ \simeq\ \frac{|a|^2 + |b|^2 + |c|^2}{|a|^2}\ |\bar x_3|\ ,
\eea
while in the case $|\bar x_3| \ll |\bar x_2| \ll |\bar x_1|$, one has:
\beq
  f_3\ \simeq\ |\bar x_1|\ ,  \qquad  f_2\ \simeq\ |\bar x_2|\ ,  \qquad
  f_1\ \simeq\ |\bar x_3|\ ,
\eeq
etc. When the hierarchy between the $\bar x_i$ is not so pronounced,
Eq. (\ref{eq:approximate_f_i}) is no longer a good approximation.
In the case $\bar x_{min} \ll \bar x_{middle} \leq \bar x_{max}$, however,
one still has:
\beq
  f_1\ \simeq\ \sqrt{C_3 / C_2}\ \sim\ \bar x_{min}\ ,
\label{eq:approximate_f_1}
\eeq
while in the case $\bar x_{min} \leq \bar x_{middle} \ll \bar x_{max}$:
\beq
  f_3\ \simeq\ \sqrt{C_1}\ \sim\ \bar x_{max}\ .
\label{eq:approximate_f_3}
\eeq

The formulae for the mixing angles are more involved, but they
simplify for some $\bar x_i$ hierarchies. It is convenient to write
$\tilde X$ as:
\bea
  \tilde X\ =\
  \bar x_3 \left( \begin{array}{ccc}
    1 & \frac{b}{a} & \frac{c}{a}  \\
    \frac{b}{a} & \frac{b^2}{a^2} & \frac{bc}{a^2}  \\
    \frac{c}{a} & \frac{bc}{a^2} & \frac{c^2}{a^2}
    \end{array}  \right)\
  +\ \bar x_2 \left( \begin{array}{ccc}
    \frac{b^2}{a^2}\, \frac{y^2_u}{y^2_c}& - \frac{b}{a}\, \frac{y_u}{y_c}
    & - \frac{b \Delta_{23}}{a \Delta_{33}}\, \frac{y_u}{y_c}  \\
    - \frac{b}{a}\, \frac{y_u}{y_c} & 1 & \frac{\Delta_{23}}{\Delta_{33}}  \\
    -  \frac{b \Delta_{23}}{a \Delta_{33}}\, \frac{y_u}{y_c}
    & \frac{\Delta_{23}}{\Delta_{33}} & \frac{\Delta^2_{23}}{\Delta^2_{33}}
    \end{array}  \right)  \nn  \\  \nn  \\
  +\ \bar x_1 \left( \begin{array}{ccc}
    \frac{\Delta^2_{13}}{\Delta^2_{33}}\, \frac{y^2_u}{y^2_t}
    & - \frac{\Delta_{13} \Delta_{23}}{\Delta^2_{33}}\, \frac{y_u y_c}{y^2_t}
    & \frac{\Delta_{13}}{\Delta_{33}}\, \frac{y_u}{y_t}  \\
    - \frac{\Delta_{13} \Delta_{23}}{\Delta^2_{33}}\, \frac{y_u y_c}{y^2_t}
    & \frac{\Delta^2_{23}}{\Delta^2_{33}}\, \frac{y^2_c}{y^2_t}
    & - \frac{\Delta_{23}}{\Delta_{33}}\, \frac{y_c}{y_t}  \\
    \frac{\Delta_{13}}{\Delta_{33}}\, \frac{y_u}{y_t}
    & - \frac{\Delta_{23}}{\Delta_{33}}\, \frac{y_c}{y_t} & 1
    \end{array}  \right)\ ,
\label{eq:X_tilde}
\eea
where subdominant terms of relative order $y_u / y_c$ and $y_c / y_t$
are understood in each entry of the matrices multiplying $\bar x_1$,
$\bar x_2$ and $\bar x_3$. Let us first consider the hierarchy
$|\bar x_1|, |\bar x_2| \ll |\bar x_3|$. In this case,
$\tilde X$ is dominated by the contribution proportional to $\bar x_3$
in Eq.~(\ref{eq:X_tilde}), and $U_{\tilde X}$ is given by:
\beq
  U_{\tilde X}\ \simeq\ \left( \begin{array}{ccc}
    \tilde a & 0 & \bar a  \\
    - \frac{\bar a^\star \bar b}{\tilde a} & \frac{\bar c^\star}{\tilde a} & \bar b  \\
    - \frac{\bar a^\star \bar c}{\tilde a} & - \frac{\bar b^\star}{\tilde a} & \bar c
    \end{array}  \right)
    \left( \begin{array}{ccc}
    a^\star_{12} & b_{12} & 0  \\
    - b^\star_{12} & a_{12} & 0 \\
    0 & 0 & 1
    \end{array}  \right) P\ ,
\eeq
where $(\bar a, \bar b, \bar c) \equiv (a, b, c) / \sqrt{|a|^2 + |b|^2 + |c|^2}$,
$\tilde a \equiv \sqrt{1 - |\bar a|^2}$, $P$ is a diagonal matrix of phases,
$|a_{12}|^2 + |b_{12}|^2 = 1$, and $a_{12}$ and $b_{12}$ depend
on the subdominant terms in $\tilde X$.
After multiplication by $U^T_q$, this gives:
\bea
  |(U_f)_{12}| & \simeq & \left| \tilde a b_{12}
    + (\bar c^\star a_{12} - \bar a^\star \bar b b_{12}) V_{cd}\,
    e^{i (\Phi^u_2 - \Phi^u_1)} / \tilde a \right|\ \gtrsim\ \lambda\ ,
    \label{eq:U12}  \\
  |(U_f)_{13}| & \simeq & \left| \bar a
    + \bar b\, V_{cd}\, e^{i (\Phi^u_2 - \Phi^u_1)} \right|\ \sim\ 1\ ,
    \label{eq:U13}  \\
  |(U_f)_{23}| & \simeq & \left| \bar b
    + \bar a\, V_{us}\, e^{-i (\Phi^u_2 -\Phi^u_1)} \right|\ \sim\ 1\ .
    \label{eq:U23}
\eea
Thus, the hierarchy $|\bar x_1|, |\bar x_2| \ll |\bar x_3|$ leads to large
right-handed neutrino mixing angles (cancellations are possible in
Eqs.~(\ref{eq:U12})-(\ref{eq:U23}) though).

Let us then consider the hierarchy $|\bar x_1|, |\bar x_3| \ll |\bar x_2|$.
In this case, depending on the value of $\bar x_3 / \bar x_2$, $\tilde X$
is dominated either by the contribution proportional to $\bar x_2$ alone,
or by both contributions proportional to $\bar x_2$ and $\bar x_3$.
$U_{\tilde X}$ is then given by:
\beq
  U_{\tilde X}\ \simeq\ \left( \begin{array}{ccc}
    1 & 0 & b_{13}  \\
    - b^\star_{13} \bar \Delta_{33} & \bar \Delta^\star_{23} & \bar \Delta_{33}  \\
    - b^\star_{13} \bar \Delta_{23} & - \bar \Delta^\star_{33} & \bar \Delta_{23}
    \end{array}  \right)
    \left( \begin{array}{ccc}
    a^\star_{12} & b_{12} & 0  \\
    - b^\star_{12} & a_{12} & 0 \\
    0 & 0 & 1
    \end{array}  \right) P\ ,
\eeq
\beq
  b_{13}\ \simeq\ \left\{ \begin{array}{lcl}
    \frac{b \bar \Delta_{33}}{a}\, \frac{y_u}{y_c}
    & & |\bar x_3| \ll \frac{y_u}{y_c}\, |\bar x_2|  \\
    \frac{(c \bar \Delta^\star_{23} + b \bar \Delta^\star_{33}) \bar \Delta^2_{33}}{a}\,
    \frac{\bar x_3}{\bar x_2}
    & & |\bar x_3| \gg \frac{y_u}{y_c}\, |\bar x_2|
  \end{array}  \right. ,
\eeq
where $(\bar \Delta_{23}, \bar \Delta_{33}) \equiv (\Delta_{23}, \Delta_{33})
/ \sqrt{|\Delta_{23}|^2 + |\Delta_{33}|^2}$, and $a_{12}$ and $b_{12}$
depend on the subdominant terms in $\tilde X$. After multiplication by
$U^T_q$, this gives:
\bea
  |(U_f)_{12}| & \simeq & \left| b_{12}
    + a_{12}\, \bar \Delta^\star_{23}\, V_{cd}\, e^{i (\Phi^u_2 - \Phi^u_1)} \right|\
    \gtrsim\ \lambda\ ,  \\
  |(U_f)_{13}| & \simeq & \left| \bar \Delta_{33}\, V_{cd}
    + b_{13}\, e^{-i (\Phi^u_2 - \Phi^u_1)} \right|\ \gtrsim\ \lambda\ ,  \\
  |(U_f)_{23}| & \simeq & \left| \bar \Delta_{33} \right|\ \sim\ 1\ .
\eea
Thus, the hierarchy $|\bar x_1|, |\bar x_3| \ll |\bar x_2|$ leads to large
right-handed neutrino mixing angles as well, but cancellations are possible
in $(U_f)_{12}$ and $(U_f)_{13}$. The same conclusion holds in
the qualitatively similar cases $|\bar x_1| \ll |\bar x_3| \lesssim |\bar x_2|$
and $|\bar x_3| \ll |\bar x_1| \lesssim |\bar x_2|$.

Let us now turn to the case $|\bar x_2|, |\bar x_3| \ll |\bar x_1|$, which
contrary to the previous ones leads to small mixing angles. In the case
of a strong ``inverted'' hierarchy $|\bar x_3| \ll |\bar x_2| \ll |\bar x_1|$,
with $|\bar x_3| \ll (y_u / y_t)\, |\bar x_1|$
and $|\bar x_2| \ll (y_c / y_t)\, |\bar x_1|$, one obtains:
\beq
U_{\tilde X}\ \simeq\ \left( \begin{array}{ccc}
    1 & \frac{b}{a}\, \frac{\bar x_3}{\bar x_2} - \frac{b}{a}\, \frac{y_u}{y_c}
    & \frac{\Delta_{13}}{\Delta_{33}}\, \frac{y_u}{y_t}  \\
    - \frac{b^\star}{a^\star}\, \frac{\bar x^\star_3}{\bar x^\star_2}
    + \frac{b^\star}{a^\star}\, \frac{y_u}{y_c}
    & 1 & - \frac{\Delta_{23}}{\Delta_{33}}\, \frac{y_c}{y_t}  \\
    - \frac{b^\star \Delta^\star_{23}}{a^\star \Delta^\star_{33}}\,
    \frac{\bar x^\star_3}{\bar x^\star_2}\, \frac{y_c}{y_t}
    + \frac{c^\star}{a^\star}\, \frac{y_u}{y_t}
    & \frac{\Delta^\star_{23}}{\Delta^\star_{33}}\, \frac{y_c}{y_t} & 1
    \end{array}  \right)\ .
\eeq
After multiplication by $U^T_q$, this gives $U_f \simeq U^T_q$. Hence
the right-handed neutrino mixing angles are given by the CKM angles,
up to corrections of order $y_u / y_c$, $y_c / y_t$ and
$\bar x_3 / \bar x_2$ (the same conclusion holds for
$|\bar x_3| \lesssim (y_u / y_t)\, |\bar x_1|$,
$|\bar x_2| \lesssim (y_c / y_t)\, |\bar x_1|$). Explicitly, one has:
\bea
  |(U_f)_{12}| & \simeq & \left| V_{cd}
    + \left( \frac{b}{a}\, \frac{\bar x_3}{\bar x_2} - \frac{b}{a}\, \frac{y_u}{y_c} \right)
    e^{-i (\Phi^u_2 - \Phi^u_1)} \right| ,  \label{eq:U12_inverted}  \\
  |(U_f)_{13}| & \simeq & \left| V_{td}
    - \frac{\Delta_{23}}{\Delta_{33}}\, \frac{y_c}{y_t}\, V_{cd}\,
    e^{i (\Phi^u_2 - \Phi^u_3)} \right| ,  \label{eq:U13_inverted}  \\
  |(U_f)_{23}| & \simeq & \left| V_{ts}
    - \frac{\Delta_{23}}{\Delta_{33}}\, \frac{y_c}{y_t}\, e^{i (\Phi^u_2 - \Phi^u_3)} \right| .
    \label{eq:U23_inverted}
\eea
The limit $|z_3|^2 \ll 4 \alpha \beta$, which yields
$\sqrt{\alpha / \beta}\, (\bar x_3, \bar x_2, \bar x_1)
= (s_3 y_u, s_2 y_c, s_1 y_t)$
($s_i = \pm\, \mbox{sign} (\mbox{Re} (z_i))$ for $x^\pm_i$),
deserves a particular discussion. In this case:
\bea
\hskip -.5 cm
U_{\tilde X}\ \simeq\ \left( \begin{array}{ccc}
    1 & (\frac{s_3}{s_2} - 1)\, \frac{b}{a} \frac{y_u}{y_c}
    & \frac{s_3 c \Delta_{33} - s_2 b \Delta_{23} + s_1 a \Delta_{13}}
    {s_1 a \Delta_{33}}\, \frac{y_u}{y_t}  \\
    - (\frac{s_3}{s_2} - 1)\, \frac{b^\star}{a^\star} \frac{y_u}{y_c}
    & 1 & (\frac{s_2}{s_1} - 1)\, \frac{\Delta_{23}}{\Delta_{33}}\, \frac{y_c}{y_t}  \\
    \left[ - (\frac{s_3}{s_2} - 1)\, \frac{c^\star}{a^\star}
    + \frac{s_3}{s_2}\, (\frac{s_2}{s_1} - 1)\,
    \frac{\Delta^\star_{13}}{\Delta^\star_{33}} \right]\! \frac{y_u}{y_t}
    & - (\frac{s_2}{s_1} - 1)\, \frac{\Delta^\star_{23}}{\Delta^\star_{33}}
    \frac{y_c}{y_t} & 1
    \end{array}  \right) .
\eea
For the two solutions characterized by $s_1 = s_2 = s_3$, cancellations
occur in the off-diagonal entries of $U_{\tilde X}$, implying
$U_{\tilde X} \simeq {\bf 1}$ and $f \simeq s_1 \sqrt{\beta / \alpha}\, Y_u$,
consistently with Eq.~(\ref{eq:f_cancel}).
For the other six solutions, one still has $U_f \simeq U^T_q$ (but with
larger corrections), hence $f \simeq \sqrt{\beta / \alpha}\,
U^T_q \mbox{Diag}\, (s_1 y_1, s_2 y_2, s_3 y_3) U_q$.
Finally, when the hierarchy is milder,
with e.g. $|\bar x_3| \gg (y_u / y_t)\, |\bar x_1|$
or $|\bar x_2| \gg (y_c / y_t)\, |\bar x_1|$, or both, $U_f$
deviates more significantly from $U^T_q$ and is characterized
by larger mixing angles (but cancellations may occur for specific values
of the ratios $\bar x_i / \bar x_j$). For instance, when
$|\bar x_3| \lesssim (y_u / y_t)\, |\bar x_1|$ and
$|V_{ts}\, \bar x_1| \ll |\bar x_2| \ll |\bar x_1|$,
one finds $|(U_f)_{12}| \simeq |V_{cd}|$,
$|(U_f)_{13}| \simeq |V_{cd}\, (\Delta_{23} / \Delta_{33})\, \bar x_2 / \bar x_1|$
and $|(U_f)_{23}| \simeq |(\Delta_{23} / \Delta_{33})\, \bar x_2 / \bar x_1|$.

With the above formulae, it is possible to explain most features of
Figs.~\ref{fig:Mi_ref} and~\ref{fig:Uij_ref}. For example, the fact that
$M_1$ takes a constant value over the considered range of values
for $v_R$ in all four solutions with $x_3 = x^-_3$ just follows from
Eq.~(\ref{eq:approximate_f_1}). As can be easily checked indeed, the conditions
$4 \alpha \beta \ll |z_3|^2$ and $|\bar x^-_3| \ll |\bar x^\pm_1|, |\bar x^\pm_2|$
are always satisfied for $\sqrt{\beta / \alpha}\, v_R \gg 10^{10}$~GeV,
implying $M_1 \simeq |\bar x^-_3| v_R \simeq m^2_u (M_{GUT}) / (a m_3)$,
namely $M_1 \approx 10^5$~GeV for $a \approx 0.2$.
Similarly, in the 2 solutions with $x_3 = x^+_3$ and $x_2 = x^-_2$,
the conditions $4 \alpha \beta \ll |z_2|^2$ and
$|\bar x^-_2| \ll |\bar x^\pm_1|, |\bar x^+_3|$ are satisfied
for $\sqrt{\beta / \alpha}\, v_R \gg 10^{12}$~GeV, implying
$M_2 \sim |\bar x^-_2| v_R \sim m^2_c (M_{GUT}) / m_3
\simeq 2 \times 10^9$~GeV.
As for the right-handed neutrino mixing angles, the fact that they
are approximately independent of $v_R$ and very close to the CKM
angles in solutions $(+,-,-)$ and $(-,-,-)$ is due to the hierarchy
$|\bar x_3| \leq (y_u / y_c)\, |\bar x_2| \leq (y_u / y_t)\, |\bar x_1|$,
which holds over the considered range of values for $v_R$.
For large values of $v_R$, both solutions have
a strong hierarchy of the $\bar x_i$'s,
$|\bar x_3| \ll (y_u / y_c)\, |\bar x_2| \ll (y_u / y_t)\, |\bar x_1|$,
and the $(U_f)_{ij}$ are given by Eqs.~(\ref{eq:U12_inverted})
to~(\ref{eq:U23_inverted}). In the other 6 solutions, one recovers
$|(U_f)_{ij}| \simeq |(V_{CKM})_{ji}|$ in the small $v_R$ region
(corresponding to the limit $|z_3|^2 \ll 4 \alpha \beta$, namely
$\sqrt{\beta / \alpha}\, v_R \ll 10^{10}$~GeV), while the large
$v_R$ region, where the hierarchy $|\bar x_2|, |\bar x_3| \ll |\bar x_1|$
is no longer satisfied, is characterized by larger values
of the right-handed neutrino mixing angles.

\end{appendix}



\end{document}